\documentclass[a4paper,11pt]{article}
\usepackage{jheppub} 
\usepackage{amsmath}
\usepackage{fontspec}
\usepackage{colortbl}
\usepackage{booktabs}
\usepackage{graphicx}
\usepackage{lineno}

\makeatletter

\providecommand{\abstract}[1]{}
\DeclareOldFontCommand{\rm}{\normalfont\rmfamily}{\mathrm}
\DeclareOldFontCommand{\sf}{\normalfont\sffamily}{\mathsf}
\DeclareOldFontCommand{\tt}{\normalfont\ttfamily}{\mathtt}
\DeclareOldFontCommand{\bf}{\normalfont\bfseries}{\mathbf}
\DeclareOldFontCommand{\it}{\normalfont\itshape}{\mathit}
\DeclareOldFontCommand{\sl}{\normalfont\slshape}{\@nomath\sl}
\DeclareOldFontCommand{\sc}{\normalfont\scshape}{\@nomath\sc}

\usepackage[nice]{nicefrac}

\author[a,b,c,d]{José Matos}
\affiliation[a]{Associate Laboratory LaPMET, 4169-007 Porto, Portugal.}
\affiliation[b]{Departamento de Física e Astronomia, Faculdade de Ciências, Universidade do Porto, rua do Campo Alegre s/n, 4169-007 Porto, Portugal.}
\affiliation[c]{Centro de Física das Universidades do Minho e Porto (CF-UM-PT), Departamento de Física e Astronomia, Faculdade de Ciências, Universidade do Porto, 4169-007 Porto, Portugal.}
\affiliation[d]{Fields and Strings Laboratory, Institute of Physics, École Polytechnique Fédérale de Lausanne (EPFL), Route de la Sorge, CH-1015 Lausanne, Switzerland} 

\emailAdd{jose.bouradematos@gmail.com}

\renewcommand{\[}{\begin{equation}}
\renewcommand{\]}{\end{equation}}

\makeatother

\abstract{
We introduce a Monte Carlo strategy for directly estimating partition function
ratios between distinct global sectors of a lattice theory. It enlarges
the configuration space to sample an interpolating family whose endpoints
are the desired sectors, and uses flat histogram methods to reconstruct
the corresponding free energy difference. Although the construction
is more general, we focus here on the three-dimensional Ising model
on the slab $\mathbb{R}^{2}\times S^{1}_{L_{z}}$ at the bulk critical
point, comparing the untwisted periodic sector with the $\mathbb{Z}_{2}$-twisted
antiperiodic sector. A large-volume and aspect ratio extrapolation
gives the symmetry-twisted thermodynamic Casimir difference $\Delta_{\mathbb{Z}_{2}}=0.327(2)$
directly, without lattice derivatives or bulk subtractions. This provides
an independent twisted sector probe of tensions observed in periodic sector
thermodynamic Casimir observables. More generally, the method gives
direct but selective numerical access to CFT compactification data,
including estimates of the effective thermal screening mass and the $\mathbb{Z}_{2}$-odd
sector energy gap on $T^{2}$.
}

\begin{document}
\title{Monte Carlo reconstruction of symmetry-twisted partition function ratios:
the critical 3D Ising}
\maketitle

\section{Summary}

Partition function ratios between distinct global sectors are finite-volume
observables of a lattice theory. At criticality they become symmetry-resolved
compactification data of the underlying CFT. When the sector is defined by a global symmetry twist, the ratio measures
the response of the theory to a background symmetry holonomy around a
non-contractible circle. Equivalently, it can be described by inserting an
invertible symmetry defect on a codimension-one surface transverse to that
circle. In this work we
study such a sector ratio in the three-dimensional Ising universality
class. We perform Monte Carlo simulations on $T^2_L\times S^1_{L_z}$, and study two limiting
channels: the slab limit $\mathbb{R}^2\times S^1_{L_z}$ and the transverse torus channel
$T^2_L\times R$. The two sectors are the untwisted or periodic sector and the
$\mathbb Z_2$-twisted or antiperiodic sector. 

The basic idea is to regard the two sectors as subsets of an enlarged
configuration space. One introduces an interpolating family of lattice
systems whose endpoint partition functions are the desired sector
partition functions, and samples this family together with the microscopic
degrees of freedom. A flat histogram bias is then tuned so that the
simulation moves efficiently between the endpoints. Once the bias
has converged, its endpoint difference gives the logarithm of the
partition function ratio, and hence the sector free energy difference.
The interpolation is part of the estimator and need not be unique.
In some cases it may also have a physical interpretation as a defect
or interface deformation.

In the Ising implementation used here, the interpolation is realized
by choosing a plane $\Sigma$ transverse to the compact direction
and assigning a variable coupling $J$ to the bonds crossing it. The
endpoint $J=1$ gives the periodic sector, while $J=-1$ gives the
antiperiodic sector, equivalently the insertion of the global $\mathbb{Z}_{2}$
symmetry operator in the trace obtained by quantizing along $S^{1}_{L_{z}}$. The simulation samples
both the Ising spins and the plane coupling $J$, and the flat histogram
reconstruction gives $F(-1)-F(1)=-\log(Z_{{\rm AP}}/Z_{{\rm P}})$.
Thus the sector difference is obtained as an endpoint free energy
difference, rather than from separate estimates of the two free energies.

We use this estimator to measure the symmetry-twisted free energy
difference in the three-dimensional Ising slab at bulk criticality,
and extract the corresponding thermodynamic Casimir difference directly,
without bulk subtractions or numerical differentiation in $L_{z}$.
Its output is nevertheless a sector difference, not an absolute thermodynamic
Casimir amplitude.  In each sector, the universal finite-volume part of the critical free
energy is
\[
f_{\alpha}(t,L_{z})=L_{z}f_{\text{bulk}}(t)+L^{-2}_{z}\Theta_{\alpha}(x),\qquad x=tL^{1/\nu}_{z},
\]
where $\alpha$ labels the boundary condition or global sector and
$\Theta_{\alpha}$ is universal \cite{Gambassi:2008beg,krechCasimirEffectCritical1994,diehlTheoryBoundaryCritical1997}.
The free energy density difference then defines a symmetry-twisted
thermodynamic Casimir difference $\Theta_{\alpha}(x)-\Theta_{\beta}(x)$. Extracting either absolute amplitude requires independent
input for the other sector.

The three-dimensional Ising universality class is the natural benchmark
for novel strategies. Monte Carlo studies, often using improved realizations
such as the Blume-Capel model, have determined $\Theta_{\alpha}(x)$
and the critical amplitudes $C_{\alpha}\equiv\Theta_{\alpha}(0)$
for several boundary conditions, including symmetry-breaking and ordinary/free
surface universality classes. Periodic systems and slab corrections
to scaling have also been studied numerically \cite{hasenbuschFiniteSizeScaling2010,bulgarelliCasimirEffectCritical2025,hasenbuschThermodynamicCasimirEffect2010,hasenbuschThermodynamicCasimirEffect2012,hasenbuschThermodynamicCasimirEffect2015,hasenbuschThermodynamicCasimirForces2013,huchtAspectratioDependenceThermodynamic2011}.

In the slab channel, \(q=L/L_z\gg 1\), the large-volume and large aspect ratio extrapolation yields the symmetry-twisted thermodynamic Casimir difference
\begin{equation}
\Delta_{\mathbb Z_2}=C_{-1}-C_1=0.327(2).
\end{equation}
At large but finite \(q\), the approach to this limit is described by transverse-channel screening states propagating across the large transverse directions. Fitting this correction gives an effective screening scale
\begin{equation}
\kappa_{\rm eff}=0.71(4).
\end{equation}
The same sector ratio also has a complementary interpretation in the opposite limit, \(q\ll 1\). Quantizing along the compact direction gives
\begin{equation}
Z_{\rm P}
=
{\rm Tr}_{\mathcal H(T_L^2)}
\left(e^{-L_z H}\right),
\qquad
Z_{\rm AP}
=
{\rm Tr}_{\mathcal H(T_L^2)}
\left(S e^{-L_z H}\right),
\end{equation}
where \(S\) implements the \(\mathbb Z_2\) symmetry. Thus the twisted sector weights each finite-volume torus state by its \(\mathbb Z_2\) parity. If the finite-volume ground state is unique and \(\mathbb Z_2\)-even, its leading contribution cancels in the sector difference, and the first nonvanishing term is controlled by the lightest \(\mathbb Z_2\)-odd state on \(T_L^2\). This gives, for the square torus, 
\begin{equation}
E_\sigma(L)-E_0(L)=\frac{\kappa_\sigma}{L}+\cdots,
\qquad
\kappa_\sigma=1.230(3).
\end{equation}
Thus the same observable gives the Casimir difference in the slab limit, while its finite aspect ratio dependence gives selective access to symmetry-resolved finite-volume spectral data.

An additional advantage of the sector difference is that it can suppress
subleading corrections that would be present in the individual sectors.
In the data analyzed here, some expected corrections are numerically
unresolved, allowing the surviving finite-volume structures to be
isolated more clearly.

Sec.~\ref{sec:theory} defines the twisted sector free energy, its
thermodynamic limit density, and the expected critical corrections.
Sec.~\ref{sec:numerics} describes the extended configuration space
and the two-level flat histogram Monte Carlo method. Sec.~\ref{subsec:2d_benchmark}
benchmarks the estimator in the critical two-dimensional Ising model.
Sec.~\ref{subsec:Three-dimensional-analysis} gives the three-dimensional
result. Sec.~\ref{subsec:discussion} discusses possible applications
and extensions. 

\section{Twisted sector free energy and scaling }\label{sec:theory}

Consider the three-dimensional Ising model on $M=\mathbb{R}^{2}_{x,y}\times S^{1}_{L_{z}}$.
A \(\mathbb Z_2\) twist in the compact direction is implemented by a
nontrivial \(\mathbb Z_2\) holonomy around \(S^1_{L_z}\). Equivalently,
one inserts the topological symmetry defect on a codimension-one surface
transverse to the compact direction. The untwisted sector is denoted by
P. On \(\mathbb R^2\times\mathbb R\) the corresponding defect plane can
be removed by a spin flip on one side. On
\(\mathbb R^2\times S^1_{L_z}\) it defines a distinct global sector.

Hamiltonian quantization along $z$ gives 
\begin{equation}
Z_{\mathrm{P}}=\mathrm{Tr}_{\mathcal{H}}\!\left(e^{-L_{z}H}\right),\qquad Z_{\mathrm{AP}}=\mathrm{Tr}_{\mathcal{H}}\!\left(Se^{-L_{z}H}\right),\label{eq:twisted-trace}
\end{equation}
where $S$ is the operator implementing the $\mathbb{Z}_{2}$ symmetry.
We define 
\begin{equation}
\Delta F_{\mathbb{Z}_{2}}(L_{z})\equiv-\log\!\left(\frac{Z_{\mathrm{AP}}}{Z_{\mathrm{P}}}\right).\label{eq:Deltaf-def}
\end{equation}
Its thermodynamic limit free energy \emph{per unit transverse area}
is denoted by 
\begin{equation}
\Delta f_{\mathbb{Z}_{2}}(L_{z})\equiv\lim_{A\to\infty}\frac{\Delta F_{\mathbb{Z}_{2}}(L_{z},A)}{A},
\end{equation}
where $A$ denotes the transverse area, $A=L^{2}$ for the square transverse torus used in the numerical simulations.

\subsection{Scaling at criticality and corrections to scaling}

Introduce a plane coupling $J$ interpolating between the endpoint
sectors: $J=1$ is untwisted and $J=-1$ is $\mathbb{Z}_{2}$ twisted.
For $F_{J}(L_{z})\equiv-\log Z_{J}(L_{z})$, differentiation with
respect to $L_{z}$ gives 
\begin{equation}
\frac{\partial F_{J}}{\partial L_{z}}=\langle H\rangle_{J}.~\label{eq:dF_dLz}
\end{equation}
On $M=\mathbb{R}^{2}\times S^{1}_{L_{z}}$, transverse translations
and rotations imply that the response to changing $L_{z}$ is governed
by the normal stress tensor, 
\begin{equation}
\langle H\rangle_{J}=A\langle T_{zz}\rangle_{J}\qquad\Longrightarrow\qquad\frac{\partial}{\partial L_{z}}\left(\frac{F_{J}}{A}\right)=\langle T_{zz}\rangle_{J}.\label{eq:f_Tzz_relation}
\end{equation}
At criticality, finite-size scaling gives the free energy per unit
area as 
\[
f(J,L_{z})=L_{z}f_{\text{bulk}}+\frac{C_{J}}{L^{2}_{z}}+\sum_{i}\frac{\alpha_{i}(J)}{L^{2+\omega_{i}}_{z}}+\sum_{n\ge1}\frac{\beta_{n}(J)}{L^{2+2n}_{z}}+\sum_{m}\frac{\gamma_{m}(J)}{L^{\widehat{\Delta}_{m}}_{z}},
\]
where $f_{\mathrm{bulk}}$ is the $J$-independent bulk free energy
density per unit volume. $C_{J}$ is the thermodynamic Casimir. The
first sum runs over bulk irrelevant operators allowed by the symmetries with RG exponents $\omega_{i}>0$.
The second contains analytic corrections allowed by the symmetries,
starting at $L^{-4}_{z}$ in $f$ (or $L^{-5}_{z}$ in $\langle T_{zz}\rangle$).
The last sum would contain defect-local irrelevant operators; for
example, a displacement operator with $\Delta=3$ \cite{Billo:2016cpy}
would contribute as $L^{-4}_{z}$ to $\langle T_{zz}\rangle$.

At the endpoint values $J=\pm1$, the continuum partition functions
are eq.~\eqref{eq:twisted-trace}, namely the untwisted and symmetry-twisted
traces of the same bulk Hamiltonian. In the continuum, this endpoint
sector is described by a global $\mathbb{Z}_{2}$ holonomy, equivalently
by an invertible topological $\mathbb{Z}_{2}$ symmetry defect, or
by a background $\mathbb{Z}_{2}$ gauge field for the ordinary $0$-form
symmetry with nontrivial holonomy along $S^{1}_{L_{z}}$ \cite{Gaiotto:2014kfa,Maeda:2025ycr}.
We therefore do not include defect-local corrections.

The sector free energy density difference then takes the form 
\begin{equation}
\Delta f_{\mathbb{Z}_{2}}(L_{z})\equiv f(-1,L_{z})-f(1,L_{z})=\frac{1}{L^{2}_{z}}\left[\Delta_{\mathbb{Z}_{2}}+\sum_{i}\delta\alpha_{i}L^{-\omega_{i}}_{z}+\sum_{n\ge1}\delta\beta_{n}L^{-2n}_{z}\right],\label{eq:DeltafZ2}
\end{equation}
where $\Delta_{\mathbb{Z}_{2}}\equiv C_{-1}-C_{1}$ is the Casimir
difference, not a conformal dimension, and $\delta\alpha_{i},\delta\beta_{n}$
denote sector differences. In the three-dimensional Ising class the
leading bulk scalar correction is
\begin{equation}
\Delta f_{\mathbb{Z}_{2}}(L_{z})=\Delta_{\mathbb{Z}_{2}}L^{-2}_{z}+\delta\alpha_{1}L^{-2-\omega_{1}}_{z}+\mathcal{O}\left(L^{-4}_{z}\right),\label{eq:DeltafZ2_ising}
\end{equation}
with $\omega_{1}=0.82951(61)$ obtained with navigator-based numerical
bootstrap \cite{Reehorst:2021hmp}. The next irrelevant operator
has $\omega_{2}=3.8956(43)$ obtained using extremal functionals \cite{Simmons-Duffin:2016wlq}.

\section{Monte Carlo method and results }\label{sec:numerics}

The numerical method is a two-level flat histogram reconstruction
of the free energy as a function of the plane coupling. It is based
on the standard ideas underlying multicanonical sampling, entropic
sampling, Wang-Landau recursion, transition matrix methods, and optimized
ensembles \cite{Berg:1991cf,Lee:1993cca,Wang:2000fzi,Trebst_2004,Viana_Lopes_2006}.
The implementation is a direct adaptation of the domain-wall method
of \cite{limaEffectiveStringTheory2025}. The new ingredient is the
use of the same endpoint free energy reconstruction at the bulk critical
point, where the endpoint difference is not a domain-wall tension
but the free energy difference between the periodic and $\mathbb{Z}_{2}$-twisted
sectors of the slab.

The MC simulations sample the Ising spins $\{s\}$ and the plane/defect
coupling $J$ jointly. Choose a fixed plane $\Sigma$ and multiply
each bond perpendicular to $\Sigma$ in the positive $\hat{n}_{\Sigma}$
direction by $J$. For intermediate $J$, the modified plane defines a local planar
deformation of the lattice action and serves as an interpolation between
the endpoint sectors. At the endpoints, $J=1$ and $J=-1$, it gives the periodic and antiperiodic
sectors, respectively. For fixed $J$ define 
\begin{equation}
e^{-F(J)}\equiv Z(J)\equiv\sum_{\{s\}}\exp\left[\beta\sum_{\langle xy\rangle}s_{x}s_{y}+\beta\left(J-1\right)\sum_{\langle x\rangle\in\Sigma}s_{x}s_{x+\hat{n}_{\Sigma}}\right],
\end{equation}
so that $e^{-F(1)}=Z_{{\rm P}}$ and $e^{-F(-1)}=Z_{{\rm AP}}$. All
three-dimensional simulations are performed at $\beta_{c}=0.221654626(5)$
\cite{Ferrenberg:2018zst}. To sample efficiently over $J$, we introduce
a bias potential $\omega(J)$ and simulate the extended configuration
space 
\[
\mathcal{Z}_{{\rm ext}}(\beta)=\int^{J_{\max}}_{J_{\min}}dJ\,e^{-F(J)+\omega(J)}.
\]
The bias is determined by a flat histogram/Wang-Landau procedure.
After Wang-Landau convergence, indicated by an approximately flat
histogram, $\omega(J)$ reproduces $F(J)$ up to an additive constant.
The endpoint difference then gives the sector free energy difference,
\[
\Delta F_{\mathbb{Z}_{2}}=F(-1)-F(1)=\omega(-1)-\omega(1).
\]

Spin configurations are updated by Wolff cluster moves. The continuous
variable $J$ is updated by a Metropolis step. At each Wang-Landau
refinement level, visiting a $J$ bin updates $\omega(J)$ by an additive
increment $\delta\omega$. We start from $\delta\omega=1$ and halve
it after each level, for 16 refinement stages. A stage ends when $(H_{\max}-H_{\min})/H_{\max}<0.05$.
After the final refinement step, the residual histogram is accumulated
and used to correct $\omega(J)$ a posteriori. The $J$ histogram
has 512 bins, with the bin number increased dynamically when needed
to keep the bias variation between neighboring bins small.

To reduce discretization effects from the binned representation of
the bias, we decompose it into a smooth part and a residual binned
contribution,
\[
\omega(J)=bJ+cJ^{2}+\Delta\omega(J).
\]
After each refinement stage, the current bias is fitted to the linear
plus quadratic form. The fitted contribution is absorbed into the
coefficients $b$ and $c$ and subtracted from the residual $\Delta\omega(J)$,
leaving the total bias $\omega(J)$ unchanged. The sampled range is
extended beyond the physical endpoints to avoid extracting $\omega(\pm1)$
from boundary bins. We use $J_{\min}=-1.05,\; J_{\max}=1.05$ and determine
$\omega(\pm1)$ from a quadratic fit to $\Delta\omega(J)$ using three
histogram points on each side of each endpoint.
\begin{figure}
\begin{centering}
\includegraphics[width=0.8\columnwidth]{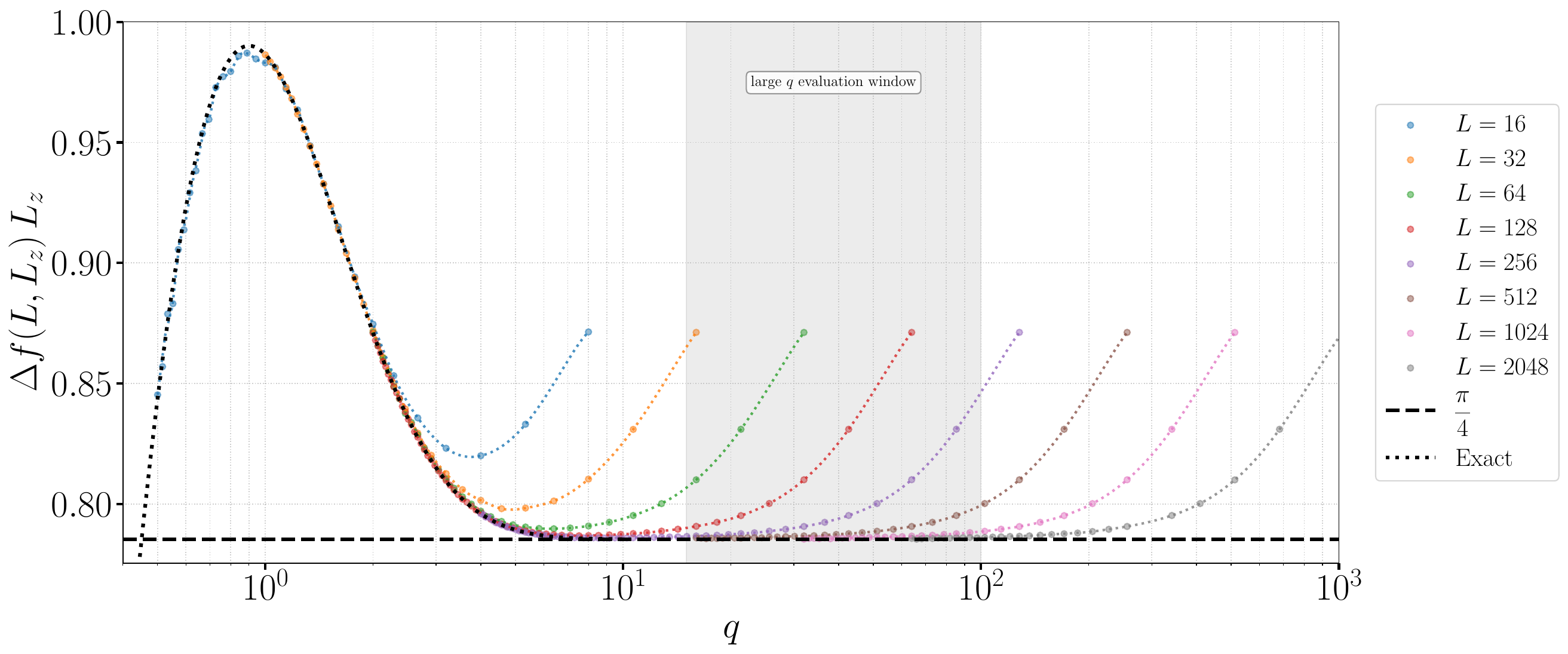} 
\par\end{centering}
\caption{Two-dimensional benchmark of the flat histogram estimator. The dimensionless
difference $L_{z}\Delta f_{\mathbb{Z}_{2}}(L,L_{z})$ is plotted against
$q=L/L_{z}$ for several transverse sizes $L$. The dashed line is
the conformal result $L_{z}\Delta f_{\mathbb{Z}_{2}}\to\pi/4$ at
$q,L_{z}\to\infty$. The black dotted curve is the exact Ising torus
result from the Jacobi-theta representation of the Virasoro characters
\cite{DiFrancesco:1997nk}. Dashed curves are cubic interpolations
in $q$ at fixed $L$. The gray band marks the large-$q$ window used
below, to study finite-$L$ effects. The deviations at small $L_{z}$/large-$q$
are the lattice and finite-volume corrections discussed in the text. }\label{fig: 2d as a function of q}
\end{figure}
\subsection{Two-dimensional benchmark }\label{subsec:2d_benchmark}

Before the three-dimensional analysis, we test the estimator in a system where
the answer is known: the critical two-dimensional Ising model on a
torus of size $L\times L_{z}$, periodic along $L$ and $\mathbb{Z}_{2}$-twisted
along $L_{z}$. The theoretical predictions for this two-dimensional
benchmark are reviewed in app.~\ref{app:2d benchmark}.

Since the transverse direction is one-dimensional, we normalize the
free energy difference per unit length, 
\begin{equation}
\Delta f_{\mathbb{Z}_{2}}(L_{z},L)\equiv\frac{F_{{\rm AP}}(L,L_{z})-F_{{\rm P}}(L,L_{z})}{L}.
\end{equation}
The critical two-dimensional theory gives, for $L\to\infty$ at fixed
$L_{z}$, 
\[
\Delta f(L_{z})=\frac{\pi}{4L_{z}},
\]
so $\Delta f(L_{z},L)L_{z}$ approaches $\pi/4$, up to finite-volume
corrections at large $q,$ which, for each $L$, corresponds to small
$L_{z}$. This expectation is tested in fig.~\ref{fig: 2d as a function of q},
where we plot $\Delta f(L_{z},L)\,L_{z}$ for transverse sizes $L=32,\ldots,2048$.
The data form a broad plateau near $\pi/4$.

\subsubsection{Finite-volume corrections }

\begin{figure}
\begin{centering}
\includegraphics[width=0.8\columnwidth]{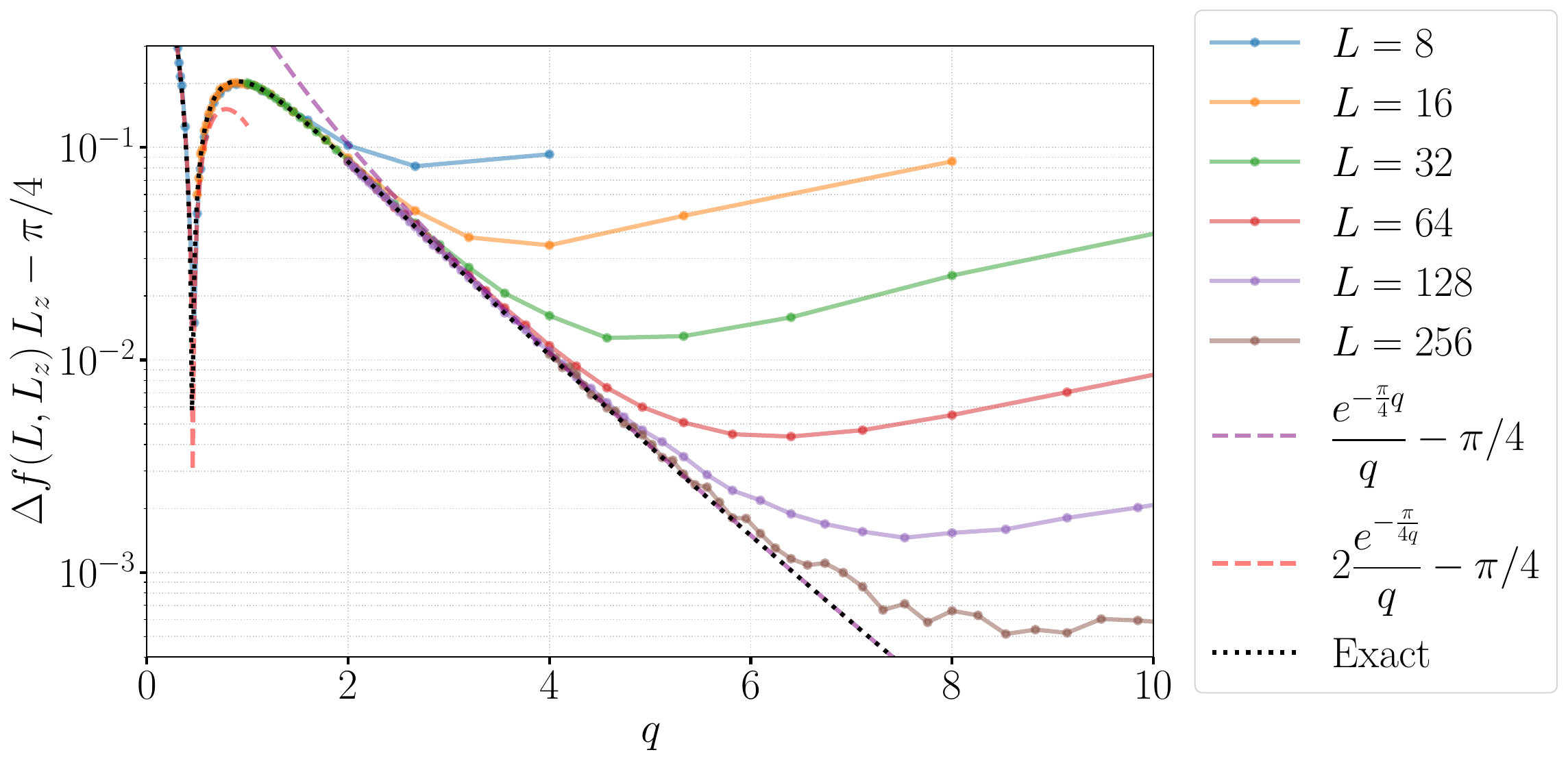} 
\par\end{centering}
\caption{Finite-$q$ corrections in the two-dimensional benchmark. The plot
shows $L_{z}\Delta f_{\mathbb{Z}_{2}}(L,L_{z})-\pi/4$ versus $q=L/L_{z}$.
The dashed curves are the leading large-$q$ and small-$q$ exponential
approximations. The exponential regime is visible only after the small-$L_{z}$
lattice corrections are controlled.  }\label{fig:exponential decay 2d}
\end{figure}

Finite-volume corrections are discussed in app.~\ref{subsec:Finite-volume-corrections}.
With $q\equiv L/L_{z}$ and $t\equiv e^{-\pi q/4}$, the double asymptotic
form is 
\begin{align*}
L_{z}\Delta f_{\mathbb{Z}_{2}}(q,L_{z})= & \frac{\pi}{4}+\frac{a_{2}}{L^{2}_{z}}+\frac{a_{4}}{L^{4}_{z}}+\mathcal{O}\left(L^{-6}_{z}\right)\\
+ & \frac{1}{q}\left[t-\frac{1}{2}t^{2}-\frac{5}{3}t^{3}\right]+\mathcal{O}\left(q^{-1}t^{4}\right)\\
+ & \mathcal{O}\left(q^{-1}tL^{-2}_{z}\right),
\end{align*}
where the $\mathcal{O}\left(q^{-1}tL^{-2}_{z}\right)$ term denotes
omitted mixed corrections and $a_{n}$ are unknown parameters.

The behavior of $L_{z}\Delta f_{\mathbb{Z}_{2}}\left(L,L_{z}\right)$
is shown in fig. \ref{fig: 2d as a function of q} and the approach to the asymptotic value in fig. \ref{fig:exponential decay 2d}. The deviations
from the asymptotic value have two origins. At fixed $L_{z}$, finite-$L$
corrections are exponentially suppressed in $L/L_{z}$. They become
visible only when $L_{z}$ is increased at fixed transverse size,
corresponding to small $q$. These are captured by the 2d Ising CFT
in finite volume, see app. \ref{app:2d benchmark}, and shown as a
dotted line in figs. \ref{fig: 2d as a function of q} and \ref{fig:exponential decay 2d}. 

\begin{figure}
\begin{centering}
\includegraphics[width=0.8\columnwidth]{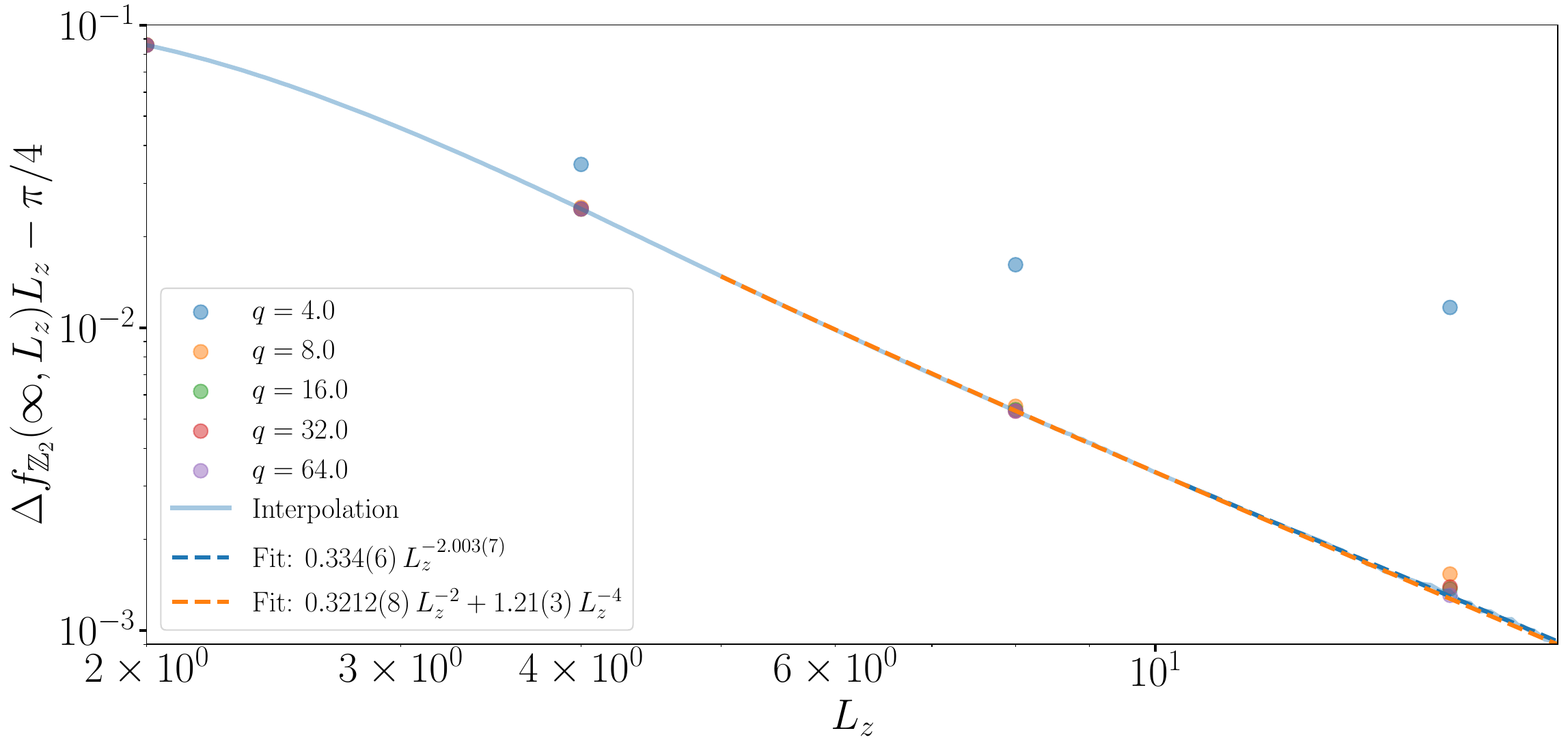} 
\par\end{centering}
\caption{Large-$q$ extrapolation at fixed $L_{z}$. For each $q\in[16,100]$,
the interpolation is evaluated and plotted versus $L_{z}$, as a single
data set. This range makes exponential finite-$q$ corrections negligible,
as indicated by fig. \ref{fig:exponential decay 2d}. Raw data at
aspect ratios with at least four values of $L_{z}$ are also shown.
The convergence from $q=4$ to $q=8$ and the stability at $q\gtrsim16$
indicate that the interpolation is in the asymptotic regime.  }\label{fig:2d as a function of Lz}
\end{figure}

At large $q$, the finite-$q$ exponential corrections are negligible.
The remaining deviations are lattice effects due to small $L_{z}$,
which occur when $L_{z}\lesssim10$. We analyze these corrections
by extrapolating to infinite aspect-ratio at fixed $L_{z}$. In practice
this means considering data with high enough $q\gtrsim16.$  Because
only a few aspect ratios are available for several transverse sizes,
we first interpolate the data in $q$ at fixed $L$. We use a cubic
interpolation to evaluate common aspect ratios. Fig. \ref{fig:2d as a function of Lz}
shows the raw Monte Carlo data together with the interpolated large-$q$
estimates for $q\in[16,100]$. We fit these estimates with a free
effective power and with the irrelevant operator ansatz described
in app. \ref{app:2d benchmark}. The free fit gives $\omega_{{\rm eff}}=2.003(7)$,
consistent with a leading $1/L^{2}_{z}$ correction to $L_{z}\Delta f_{\mathbb{Z}_{2}}$.
This is the scaling expected from the level-four irrelevant deformations
of the isotropic square-lattice Ising model, $T^{2}+\bar{T}^{2}$
and possibly $T\bar{T}$, although the latter is conjectured to be
absent at criticality, at least for the nearest-neighbor square-lattice
\cite{Caselle:2001jv}.

\subsection{Three-dimensional analysis }\label{subsec:Three-dimensional-analysis}

The three-dimensional measurement gives the sector free energy difference
per unit transverse area at finite transverse size, $\Delta f_{\mathbb{Z}_{2}}(L,L_{z})$.
We analyze the dimensionless combination 
\begin{equation}
Y(L,L_{z})\equiv L^{2}_{z}\Delta f_{\mathbb{Z}_{2}}(L,L_{z}),\qquad q\equiv\frac{L}{L_{z}}.\label{eq:3d_Y_definition}
\end{equation}

The slab, $q\gg1$, Casimir amplitude is obtained by taking both the
infinite volume and aspect ratio limit. The two numerical orders of
limits are expected to give the same result
\begin{equation}
\Delta_{\mathbb{Z}_{2}}=\lim_{q\to\infty}\lim_{L_{z}\to\infty}Y(qL_{z},L_{z})=\lim_{L_{z}\to\infty}\lim_{q\to\infty}Y(qL_{z},L_{z}).\label{eq:DeltaZ2_limits}
\end{equation}

We use the large-$q$ limit at fixed $L_{z}$ as a diagnostic. In
this regime the available data are close to the slab limit, and the
remaining deviations are finite-$L_{z}$ power corrections. This analysis
identifies the correction terms used in the subsequent fixed-$q$
infinite-volume extrapolation. The final estimate of $\Delta_{\mathbb{Z}_{2}}$
is obtained by first extrapolating to infinite volume at fixed $q$,
to obtain $Y_{\infty}(q)$, and then studying the large-$q$ behavior.
Taking the limits in this order is more stable because the approach
of $Y_{\infty}(q)$ to the slab regime is exponentially suppressed
in $q$, whereas large-$q$ data at finite $L_{z}$ approach the same
limit with power corrections in $L_{z}$.

\begin{figure}
\begin{centering}
\includegraphics[width=0.8\columnwidth]{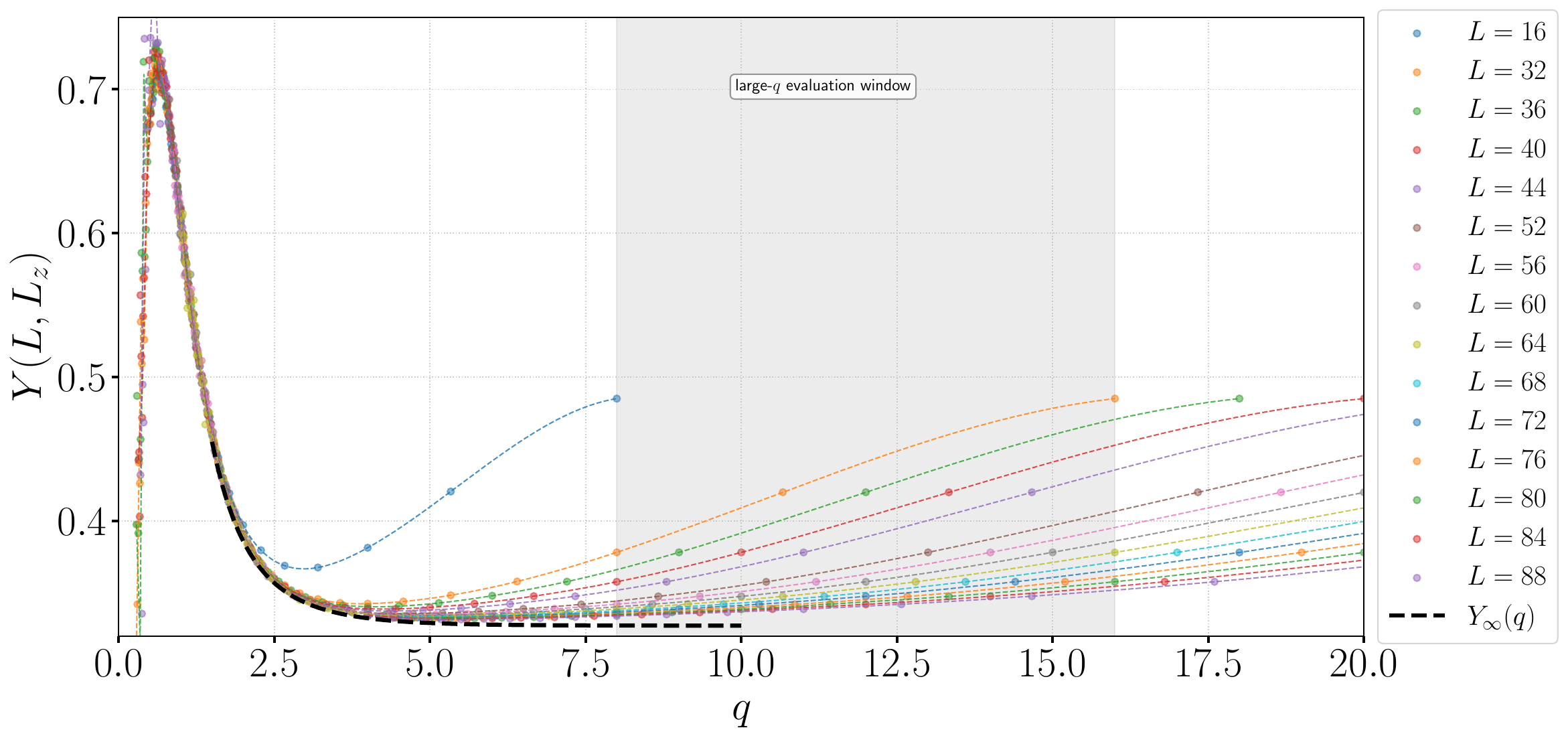} 
\par\end{centering}
\caption{Raw finite-volume data for the three-dimensional slab at $\beta=\beta_{c}=0.221654626(5)$.
The plot shows $Y(L,L_{z})=L^{2}_{z}\Delta f_{\mathbb{Z}_{2}}(L,L_{z})$
versus $q=L/L_{z}$ for several transverse sizes $L$. Dashed curves
are cubic interpolations in $q$ at fixed $L$. The dashed black line
is the infinite volume extrapolation. The gray band marks the large-$q$
window used for the fixed-$L_{z}$ diagnostic in fig. \ref{fig:extrapolation Lz}.
Unlike the two dimensional case, there is no clear plateau. }
\label{fig:3d_raw} 
\end{figure}

The raw data are shown in fig. \ref{fig:3d_raw}. Unlike the two-dimensional
case, finite-size effects are too large for the Casimir amplitude
to be read off directly from a plateau. For each fixed transverse
size, the data are interpolated as a function of $q$ \footnote{Repeating the full analysis with linear interpolation in $q$ shifts
the final value of $\Delta_{\mathbb{Z}_{2}}$ and $\kappa$ by amounts
smaller than the quoted uncertainties.}. The interpolation compares different sizes at fixed aspect ratio
and evaluates the large-$q$ regime at fixed $L_{z}$. Fig.~\ref{fig:3d_raw}
omits pointwise error bars. The final uncertainty is estimated from
the stability under fit windows and extrapolation ansätze, using the
dense set of simulated volumes.

\begin{figure}
\begin{centering}
\includegraphics[width=0.8\columnwidth]{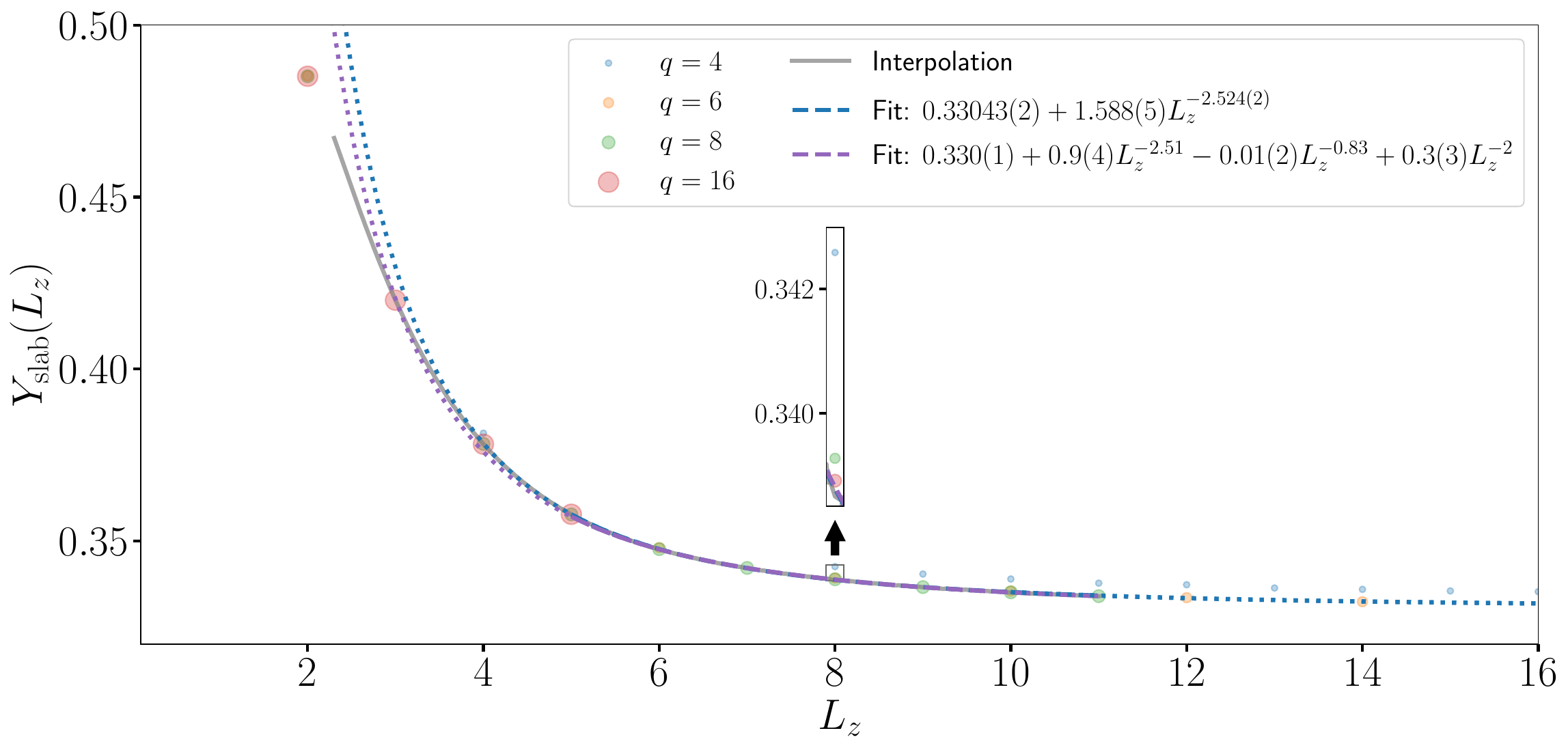} 
\par\end{centering}
\caption{Large-$q$ evaluation of the three-dimensional slab data at fixed
$L_{z}$. For each $L_{z}$, the interpolation is evaluated at $q=L/L_{z}\in[8,16]$,
where finite-$q$ corrections are not visible within the statistical
precision. The values are plotted versus $L_{z}$ and show the finite-$L_{z}$
approach to the slab Casimir amplitude. Raw data at fixed aspect ratios
with several $L_{z}$ values are also shown. A free power-law fit
gives an effective exponent near $2.5$; the second fit includes terms
proportional to $L^{-(\Delta_{T^{\prime}}-3)}_{z}$, $L^{-(\Delta_{\epsilon^{\prime}}-3)}_{z}$,
and $L^{-2}_{z}$. The uncertainties quoted in the fit labels are
formal errors from the fit covariance matrix only; they do not include
correlations, interpolation effects, or the systematic uncertainty
from the extrapolation ansatz.  }\label{fig:extrapolation Lz}
\end{figure}

As in the two-dimensional benchmark, we use the large-$q$ data at
fixed $L_{z}$ as a diagnostic of the slab regime. For each $L_{z}$,
the interpolation is evaluated in the window $q\in[8,16]$. In this
range the estimates are stable under changes of $q$ at fixed $L_{z}$,
and the available raw data at fixed aspect ratios are consistent with
the interpolated values. Fig.~\ref{fig:extrapolation Lz} shows the
resulting large-$q$ estimates as a function of $L_{z}$. This diagnostic
is used to identify the correction powers that enter the subsequent
fixed-$q$ extrapolation.

The resulting large-$q$ values are fitted as a function of $L_{z}$.
We compare two fit forms, 
\begin{align}
Y_{{\rm slab}}(L_{z}) & =\Delta_{\mathbb{Z}_{2}}+aL^{-\omega_{{\rm eff}}}_{z},\label{eq:ansatz 1}\\
Y_{{\rm slab}}(L_{z}) & =\Delta_{\mathbb{Z}_{2}}+a_{T^{\prime}}L^{-(\Delta_{T^{\prime}}-3)}_{z}+a_{\epsilon^{\prime}}L^{-(\Delta_{\epsilon'}-3)}_{z}+bL^{-2}_{z}\label{eq:ansatz22}
\end{align}
 The first fit is diagnostic. It determines the effective power visible
over the available range. It gives $\omega_{{\rm eff}}\simeq2.5$,
close to the exponent of the leading nonconserved spin-two primary
$T^{\prime}$. This does not imply that lower powers are forbidden.
Rather, it indicates that the $\epsilon^{\prime}$ and $L^{-2}_{z}$
sectors have small effective amplitudes in the sector difference over
the fitted range. App. \ref{app:3d_corrections} discusses the allowed
operators.

To understand why the scalar correction may be suppressed, it is useful
to consider the related geometry $S^{2}\times S^{1}$. In this case,
radial quantization expresses the finite-volume trace directly in
terms of CFT data; see app. \ref{par:Difference-one-pt torus}. The
argument based on radial quantization does not determine the torus
amplitudes relevant for the simulations, but it illustrates a general
mechanism. In a sector difference, the scalar correction is controlled
by the difference of scalar one-point coefficients, not by either
coefficient separately. Thus, even if
\begin{equation}
L^{\Delta_{\epsilon'}}_{z}\langle\epsilon'\rangle_{s}\neq0,\qquad s={\rm AP},{\rm P},
\end{equation}
the sector difference
\begin{equation}
L^{\Delta_{\epsilon'}}_{z}\left(\langle\epsilon'\rangle_{{\rm AP}}-\langle\epsilon'\rangle_{{\rm P}}\right)
\end{equation}
can be small. In the trace interpretation, this suppression occurs
because the leading vacuum contribution is common to the two sectors,
while the difference is sensitive only to subleading parity-weighted
states. The scalar correction proportional to $L^{-(\Delta_{\epsilon'}-3)}_{z}$
can therefore be smaller than a naive estimate based on the individual
one-point functions would suggest.

By contrast, the leading nonconserved spin-two primary can contribute through the
slab geometry scalar
\[
Q^{\mu\nu}T'_{\mu\nu},\qquad
Q_{\mu\nu}=n_\mu n_\nu-\frac{1}{3}\delta_{\mu\nu}.
\]
It gives a correction proportional to \(L_z^{-(\Delta_{T'}-3)}\) in
\(Y=L_z^2\Delta f_{\mathbb Z_2}\). There is no corresponding trace level reason
for this spin-two sector difference to be small. The same geometric selection rule
also allows the leading even spin-four primary \(C_{\mu\nu\rho\sigma}\), contracted
with the symmetric-traceless part of \(n_\mu n_\nu n_\rho n_\sigma\). Since
\(\Delta_C-3\simeq 2.02\), this correction is asymptotically larger than the
\(T'\) correction. In practice it is difficult to separate this contribution from
analytic \(L_z^{-2}\) corrections over the available range.

The second fit includes the leading correction sectors allowed by
the three-dimensional Ising CFT and by the slab geometry: the scalar
$\epsilon^{\prime}$ sector, the nonconserved spin-two sector, and
an effective $L^{-2}_{z}$ sector absorbing analytic and spin-four
contributions. The fit amplitudes are highly correlated over the available
$L_{z}$ range. Thus, they do not yield a reliable determination of
the subleading amplitudes. 

The two fits are used to estimate the dependence of $\Delta_{\mathbb{Z}_{2}}$
on the choice of ansatz. Based on the results in fig. \ref{fig:extrapolation Lz},
we use the two extrapolation ansätze to obtain $Y_{\infty}(q)$. The spread between the two extrapolations is used as the systematic uncertainty
associated with the finite-$L_z$ ansatz.

\begin{figure}
\begin{centering}
\includegraphics[width=0.8\columnwidth]{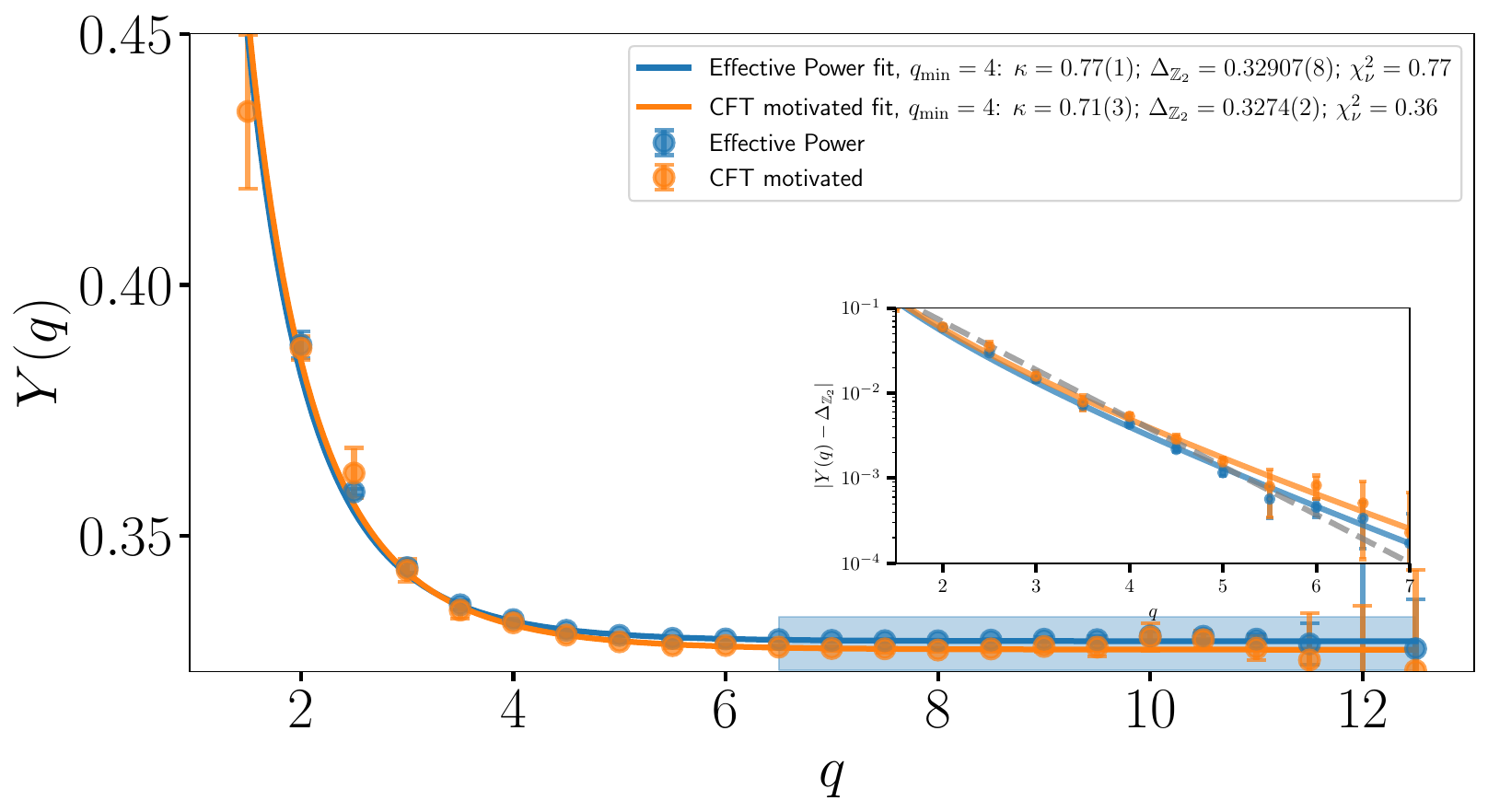} 
\par\end{centering}
\caption{Large-$q$ approach to the slab Casimir amplitude. The data are the
large-volume estimates $Y_{\infty}(q)$ obtained at fixed aspect ratio
after finite-$L_{z}$ extrapolation. The horizontal asymptote is $\Delta_{\mathbb{Z}_{2}}=\lim_{q\to\infty}Y_{\infty}(q)$.
The two curves correspond to the two finite-$L_{z}$ extrapolation
procedures and are fitted with the large-$q$ correction in eq. \eqref{eq:leading 1pt correction}.
The inset subtracts the fitted $\Delta_{\mathbb{Z}_{2}}$. The spread
between the two extrapolations is included in the systematic uncertainty.
The fit-label errors are those of the representative $q_{\min}=4$
fits. The final uncertainties are estimated from the $q_{\min}$ scan
and from the spread between the two finite-$L_{z}$ extrapolation
procedures, as described in the text.  }\label{fig:extrapolation}
\end{figure}

The infinite volume extrapolations give the curves $Y_{\infty}(q)$
in fig. \ref{fig:extrapolation}. The $q$ interpolation is used only
to compare transverse sizes at common aspect ratios. Interpolated
points are not treated as independent data and are evaluated on a
grid comparable in density to the original measurements.

We model the remaining corrections to the slab limit by the lightest
screening state in the transverse channel, as discussed in app. \ref{subsec:one_particle_3d}.
The large-$q$ fit uses 
\begin{equation}
Y_{\infty}(q)=\Delta_{\mathbb{Z}_{2}}+2\sqrt{\frac{\kappa}{2\pi}}q^{-3/2}e^{-\kappa q}+\cdots,\label{eq:leading 1pt correction}
\end{equation}
where $\kappa$ is an effective dimensionless screening mass. It is
expected to coincide with the lightest thermal screening mass that
contributes to the sector difference, but an independent spectral
measurement is needed for a direct identification. The factor $q^{-3/2}e^{-\kappa q}$
is the one-particle transverse-channel correction derived in app.
\ref{subsec:one_particle_3d}. Assuming no symmetry-protected degeneracy
for the finite-volume ground state or the lightest screening state
fixes the degeneracy factor in eq. \eqref{eq:leading 1pt correction}.
As a check, we replace the fixed coefficient $2\sqrt{\kappa/(2\pi)}$ in eq.~\eqref{eq:leading 1pt correction} by a free amplitude. The fitted amplitude is compatible with the fixed normalization and leaves $\kappa$ unchanged within the quoted uncertainty.

\begin{figure}[t]
\centering %
\begin{minipage}[t]{0.39\textwidth}%
\centering \vspace{0pt}
 \includegraphics[width=1\columnwidth]{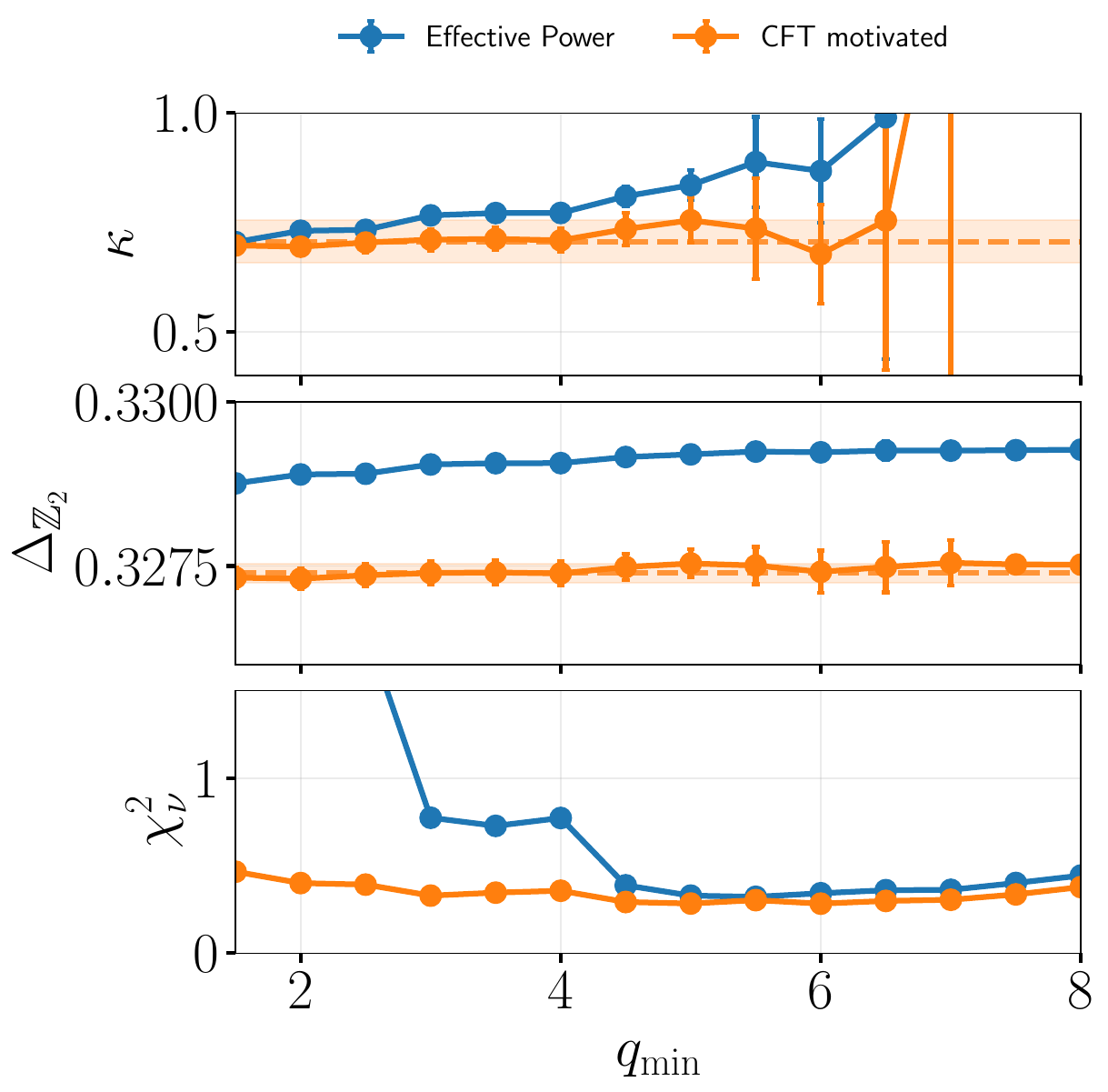} %
\end{minipage}\hfill{}%
\begin{minipage}[t]{0.6\textwidth}%
\centering{\small\setlength{\tabcolsep}{1.2pt} 
\global\long\def\arraystretch{0.95}%
 }

{\small \setlength{\tabcolsep}{2.5pt} \renewcommand{\arraystretch}{0.95} \begin{tabular}[t]{@{}c|c|c|c|c|c|c@{}} $q_{\min}$ & $\kappa$ (eff) & $\kappa$ (CFT) & $\Delta_{\mathbb{Z}_{2}}$ (eff) & $\Delta_{\mathbb{Z}_{2}}$ (CFT) & $\chi^{2}_{\nu}$ (eff) & $\chi^{2}_{\nu}$ (CFT) \\ \midrule
$1.5$ & $0.697(6)$ & $0.69(2)$ & $0.32871(6)$ & $0.3273(2)$ & $4.1$ & $0.43$ \\
$2$ & $0.731(8)$ & $0.69(2)$ & $0.32889(7)$ & $0.3273(2)$ & $1.8$ & $0.4$ \\
$2.5$ & $0.733(9)$ & $0.70(2)$ & $0.32890(7)$ & $0.3274(2)$ & $1.9$ & $0.39$ \\
$3$ & $0.77(1)$ & $0.71(2)$ & $0.32905(7)$ & $0.3274(2)$ & $0.77$ & $0.33$ \\
$3.5$ & $0.77(1)$ & $0.71(3)$ & $0.32907(8)$ & $0.3274(2)$ & $0.73$ & $0.35$ \\
$4$ & $0.77(1)$ & $0.71(3)$ & $0.32907(8)$ & $0.3274(2)$ & $0.77$ & $0.36$ \\
$4.5$ & $0.81(2)$ & $0.73(4)$ & $0.32916(8)$ & $0.3275(2)$ & $0.39$ & $0.29$ \\
$5$ & $0.84(3)$ & $0.75(5)$ & $0.32920(9)$ & $0.3275(2)$ & $0.33$ & $0.28$ \\
$5.5$ & $0.9(1)$ & $0.7(1)$ & $0.3292(1)$ & $0.3275(3)$ & $0.32$ & $0.3$ \\
$6$ & $0.9(1)$ & $0.7(1)$ & $0.3292(1)$ & $0.3274(3)$ & $0.34$ & $0.28$ \\ \end{tabular} 
}
\end{minipage}

\caption{Stability of the final large-$q$ extrapolation under variation of
the lower fit cutoff $q_{\min}$. Left: effective thermal screening
mass $\kappa$, Casimir difference $\Delta_{\mathbb{Z}_{2}}$, and
reduced chi-square $\chi^{2}_{\nu}$ for the two finite-$L_{z}$ extrapolation
procedures used to construct $Y_{\infty}(q)$. Right: numerical values
in the stable fit range. The labels \textquotedblleft eff\textquotedblright{}
and \textquotedblleft CFT\textquotedblright{} denote the effective power
and CFT-motivated finite-$L_{z}$ extrapolations. The difference between
the two procedures is used as the systematic uncertainty. The shaded
regions show the fit uncertainties of the central values quoted in
the text. }
\label{fig:qmin-stability} 
\end{figure}

The CFT-motivated large-$q$ fit is stable under changes of the lower
cutoff $q_{\min}$. Fig. \ref{fig:qmin-stability} shows weak variation
of $\Delta_{\mathbb{Z}_{2}}$ over $q_{\min}\in[1.5,8]$ \footnote{For larger $q_{\text{min}}$ the fits become unstable because the
size of the fitted corrections approach the statistical error.}. The reduced $\chi^{2}_{\nu}$ is nearly constant and below one,
$\chi^{2}_{\nu}\simeq0.3$-$0.5$. The stable values $\chi^{2}_{\nu}<1$
suggest that the fit describes the data well and that the uncertainties
propagated from the fixed-$q$ extrapolation are conservative.

We take the CFT-motivated extrapolation as the central determination.
Averaging its values over $q_{\min}\in[1.5,6]$ gives 
\begin{equation}
\Delta_{\mathbb{Z}_{2}}=0.3274(1)_{\text{fit window}},
\end{equation}
where, as the residual fit window uncertainty, we assigned half of
the spread over this range, 
\begin{equation}
\delta_{q_{\min}}\Delta_{\mathbb{Z}_{2}}=\frac{1}{2}\left[\max_{q_{\min}\in[1.5,6]}\Delta_{\mathbb{Z}_{2}}-\min_{q_{\min}\in[1.5,6]}\Delta_{\mathbb{Z}_{2}}\right]=0.00012.
\end{equation}
The dominant uncertainty comes from the finite-$L_{z}$ extrapolation
ansatz. We estimate it from the difference between the CFT-motivated
and effective power averages over the same $q_{\min}$ range, 
\begin{equation}
\delta_{{\rm ansatz}}\Delta_{\mathbb{Z}_{2}}=\left|\overline{\Delta}_{{\rm eff}}-\overline{\Delta}_{{\rm CFT}}\right|\sim0.0017.
\end{equation}
Thus our final result is 
\begin{equation}
\Delta_{\mathbb{Z}_{2}}=0.3274(1)_{q_{\min}}(17)_{{\rm ansatz}}.\label{eq:DeltaZ2_separated_errors}
\end{equation}
Combining the uncertainties into one conservative error bar gives
\begin{equation}
\Delta_{\mathbb{Z}_{2}}=0.327(2),\label{eq:DeltaZ2_final}
\end{equation}
consistent with the fixed-$L_{z}$ large-$q$ diagnostic.

For the effective screening mass, only the CFT-motivated extrapolation
is stable under changes of $q_{\text{min}}.$ Averaging over $q_{\min}\in[1.5,6]$
and assigning half of the spread as the fit window uncertainty gives
\begin{equation}
\kappa_{{\rm eff}}=0.71(4).\label{eq:kappa_eff}
\end{equation}
It is expected to coincide with the lightest thermal screening mass
that contributes to the sector difference, but an independent spectral
measurement is needed for a direct identification. The quoted uncertainties
come from changes of fit window and extrapolation ansatz.

\paragraph*{Small-$q$ channel\protect \\
}

The raw data also test the opposite channel, $q=L/L_{z}\ll1$. Whereas
the large-$q$ regime is controlled by the thermal screening mass,
the small-$q$ regime treats the compact direction as the long transfer matrix
direction. The spectrum is that of the critical theory quantized on
$T^{2}_{L}.$

The $\mathbb{Z}_{2}$ twist weights each finite-volume state by its
$\mathbb{Z}_{2}$ parity. If the finite-volume ground state is unique
and $\mathbb{Z}_{2}$ even, the leading ground state contribution
cancels in the sector difference. The first non-cancelling term is
then the lightest $\mathbb{Z}_{2}$-odd state on the transverse torus.
Thus
\[
E_{\sigma}(L)-E_{0}(L)=\frac{\kappa_{\sigma}}{L}+\cdots,
\]
and
\begin{equation}
Y\left(qL_{z},L_{z}\right)\sim\frac{A_{\sigma}}{q^{2}}e^{-\kappa_{\sigma}/q}\label{eq:small_q_asymptotic}
\end{equation}

The amplitude $A_{\sigma}$ has a spectral interpretation. If the
finite-volume ground state has degeneracy $g_{0}$ and the lightest
odd level has degeneracy $g_{\sigma}$ 
\begin{equation}
A_{\sigma}=2\,\frac{g_{\sigma}}{g_{0}}.
\end{equation}

\begin{figure}
\begin{centering}
\includegraphics[width=0.8\columnwidth]{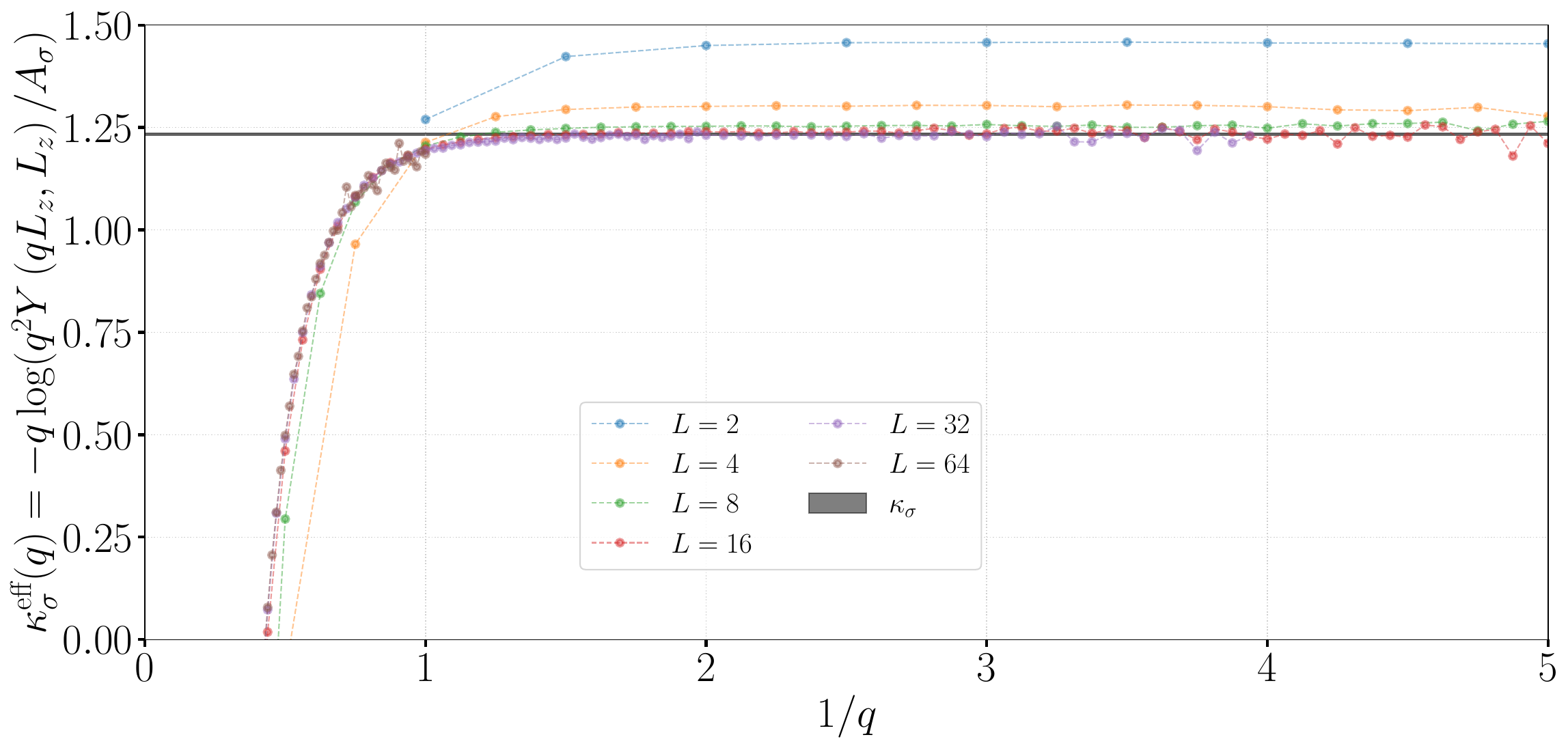} 
\par\end{centering}
\caption{Small-$q$ effective exponent in the three-dimensional twisted sector
free energy difference. The local estimator $\kappa^{{\rm eff}}_{\sigma}(q)$,
defined in eq.~\eqref{eq:kappa-sigma-eff}, with $A_{\sigma}=2$, is
plotted versus $1/q$. Data from different transverse sizes collapse
onto a common plateau. The horizontal line is the fitted value $\kappa_{\sigma}=1.230(3)$,
shown in fig. \ref{fig:small-q-stability}.}
\label{fig:small-q-effective} 
\end{figure}

For the finite-volume geometry considered here, the ground state is
expected to be unique, $g_{0}=1$, and the lowest $\mathbb{Z}_{2}$-odd
state is expected to be nondegenerate. Hence $A_\sigma=2$, with no
symmetry-imposed degeneracy factor. We use this value to define the
effective local estimator of the energy gap
\begin{equation}
\kappa^{{\rm eff}}_{\sigma}(q)=-q\log\left[\frac{q^{2}Y(qL_{z},L_{z})}{A_{\sigma}}\right].\label{eq:kappa-sigma-eff}
\end{equation}

\begin{figure}
\begin{centering}
\includegraphics[width=0.8\columnwidth]{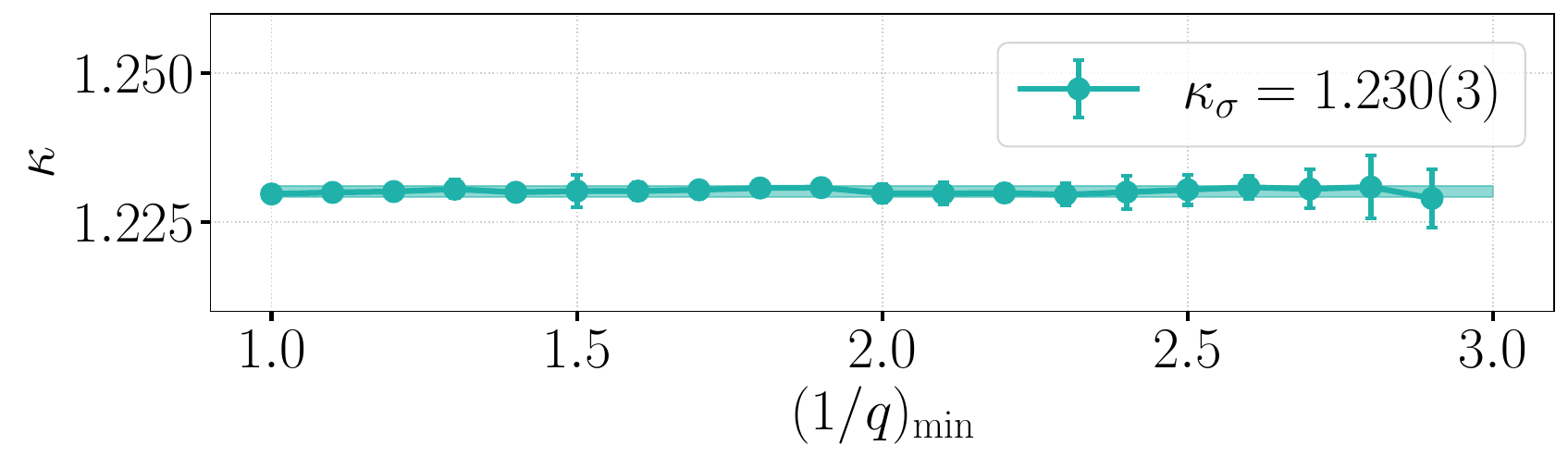} 
\par\end{centering}
\caption{Stability of the small-$q$ exponent in eq. \eqref{eq:small_q_asymptotic},
with $A_{\sigma}=2$, under variation of the lower fit cutoff $(1/q)_{\min}$, for $L=32$.
 }\label{fig:small-q-stability}
\end{figure}

If the small-$q$ channel is dominated by a single $\mathbb{Z}_{2}$-odd
state, $\kappa^{{\rm eff}}_{\sigma}(q)$ should approach a constant
as $1/q\to\infty$. This is shown in Fig. \ref{fig:small-q-effective},
where the plateau is reached already for moderate values of $1/q$. 
Letting $A_{\sigma}$ vary gives a value statistically compatible
with 2. The deviations at small $1/q$ are due to subleading odd states.
We therefore extract the $\mathbb{Z}_{2}$-odd sector gap in the small-$q$
channel by fitting
\begin{equation}
q^{2}Y(qL_{z},L_{z})=2e^{-\kappa_{\sigma}/q}+A_{\sigma^{\prime}}e^{-\kappa_{\sigma^{\prime}}/q}
\end{equation}
over several lower cutoffs in $1/q$, where $\kappa_{\sigma^{\prime}}$
is the next $\mathbb{Z}_{2}$-odd energy gap and $A_{\sigma^{\prime}}$
its degeneracy. With the current precision, the subleading parameters
vary too widely with $(1/q)_{\text{min}}$ for us to quote a meaningful
result. Fig. \ref{fig:small-q-stability} shows the stability of $\kappa_{\sigma}$
under changes of $\left(1/q\right)_{\text{min}}$. Averaging over
the accepted range and taking half the variation over the averaging
window as an error, we obtain
\begin{equation}
\kappa_{\sigma}=1.230(3).
\end{equation}

\section{Conclusion}\label{subsec:discussion}

The main result is a direct determination of a symmetry-resolved finite-volume observable of the critical three-dimensional Ising CFT. We compute the free energy difference between the periodic and $\mathbb Z_2$-twisted sectors on $T^{2}_{L}\times S^{1}_{L_z}$ by reconstructing the endpoint difference in an enlarged configuration space. The resulting estimator gives the sector difference directly, without separate free energy measurements, bulk subtractions, or numerical derivatives in $L_z$.

In the infinite-volume slab limit we find the symmetry-twisted thermodynamic
Casimir difference
\[
\Delta_{\mathbb{Z}_{2}}=0.327(2).
\]
Together with an independent determination of the periodic sector
amplitude, this gives $C_{\text{AP}}=C_{\text{P}}+\Delta_{\mathbb{Z}_{2}}.$
The coefficient $\Delta_{\mathbb{Z}_{2}}$ is the sector difference
of the leading finite-size stress tensor response on $\mathbb{R}^{2}\times S^{1}_{L_{z}}$.
It is therefore a universal compactification datum of the Ising CFT,
resolved by the $\mathbb{Z}_{2}$ holonomy around the compact direction.

The finite aspect ratio dependence gives additional spectral information.
In the large-$q$ channel, the approach to the slab limit is compatible
with the one-particle transverse correction with an effective screening scale
\[
\kappa_{{\rm eff}}=0.71(4).
\]
This scale is expected to be related to the lightest thermal screening
mass, although an independent spectral determination is required for
a direct identification. In the small-$q$ channel, the compact direction
is the long transfer matrix direction. The sector difference cancels
the $\mathbb{Z}_{2}$-even ground state contribution and isolates
the lightest $\mathbb{Z}_{2}$-odd state on the transverse torus.
This gives
\[
\kappa_{\sigma}=1.230(3).
\]
This value is consistent with the spin-channel correlation length
amplitude in \cite{PhysRevB.62.6343}, which gives
\(\kappa_\sigma=1.222(5)\) after converting \(\xi_\sigma/L\) to
\(L/\xi_\sigma\). It is also close to the speed of light normalized
Hamiltonian torus gap measured by \cite{PhysRevLett.117.210401}, which finds
\(\kappa_\sigma\simeq 1.28\) for the square torus. The latter comparison
is less direct since it involves a different microscopic realization and
an independent determination of the emergent speed.

The observable studied here is one example of a broader class of symmetry-resolved finite-volume partition functions. The flat histogram construction requires only an interpolating configuration space whose endpoints are the desired global sectors, and is therefore not specific to the three-dimensional Ising model or to the square transverse torus used in the numerical analysis.

One direct extension is to vary the compactification geometry. Changing the aspect ratio and shear of the transverse torus would give the moduli dependence of the symmetry-twisted partition function. In the slab channel, this probes shape-dependent compactification data. In the opposite channel, it gives the dependence of the symmetry-resolved finite-volume spectrum on the transverse geometry. 

The same endpoint reconstruction strategy can be applied to other global symmetries. For an $O(N)$-invariant lattice action, a twist by $g\in O(N)$ along the compact direction defines
\[
Z_g=\mathrm{Tr}_{\mathcal H(T^2_L)}\left(U_g e^{-L_z H}\right).
\]
On the lattice, this can be implemented by inserting a cut plane $\Sigma$ across which the coupling is rotated,
\[
\phi_x^a\phi_{x+\hat n_\Sigma}^a
\;\longrightarrow\;
\phi_x^a g_{ab}\phi_{x+\hat n_\Sigma}^b .
\]
Analogous constructions apply to center symmetry sectors and, more generally, to higher-form symmetry sectors in lattice gauge theory. In each case, the practical usefulness of the method depends on the existence of an efficient interpolating ensemble.

A further application is to use the same interpolation as a probe of planar defects in other lattice models. In the present work, the intermediate values of $J$ serve mainly as an efficient route between the two endpoint sectors. In models where the interpolating plane coupling has an independent physical interpretation, however, the reconstructed $J$-dependent free energy can be used to study defect deformations. This may provide information complementary to correlator-based defect spectroscopy, including possible defect RG flows and defect fixed points.

More generally, symmetry-twisted partition functions measure the response
of a theory to background symmetry fields \cite{Gaiotto:2014kfa,Maeda:2025ycr}.
Away from criticality, such observables can diagnose the sector structure
of phases with global symmetry \cite{Maeda:2025ycr}. They can also be used
as numerical probes of critical phenomena \cite{Akiyama:2026dzg}. At criticality,
the corresponding finite-volume quantities define universal compactification
data of the CFT.  The endpoint reconstruction method developed here provides a direct numerical tool for testing this framework in lattice systems.

\acknowledgments I thank A. Antunes, M. Caselle, J. Penedones, and
J. Viana for useful discussions. This work was supported by FCT -
Fundação para a Ciência e Tecnologia, I.P., through doctoral grant
2021.04743.BD \href{https://doi.org/10.54499/2021.04743.BD}{https://doi.org/10.54499/2021.04743.BD}.
I acknowledge computing time provided by INCD, funded by FCT and FEDER
under grants 2021.09830.CPCA, 2023.11029.CPCA, and 2024.09383.CPCA,
as well as GRID FEUP. I also thank the Centro de Física do Porto,
funded by the Portuguese Foundation for Science and Technology (FCT)
through Strategic Funding UIDB/04650/2020.

\appendix

\section{Two-dimensional Ising }

\label{app:2d benchmark}

In two dimensions the $\mathbb{Z}_{2}$-twisted thermodynamic Casimir
difference is known exactly. See \cite{DiFrancesco:1997nk,Ginsparg:1988ui,Aasen:2016dop}
for details. Consider the Ising CFT on a rectangular torus $T^{2}=S^{1}_{L}\times S^{1}_{L_{z}}$,
with $L$ the length of the defect and $L_{z}$ the thermal circle
size. Set 
\[
\tau=i\frac{L}{L_{z}},\qquad q=e^{2\pi i\tau}=e^{-2\pi L/L_{z}}.
\]

For a non-anomalous $\mathbb{Z}_{2}$ symmetry, two equivalent torus
amplitudes are
\[
Z^{(\mathbb{Z}_{2})}(\tau,\bar{\tau})=\mathrm{Tr}_{\mathcal{H}}\left(Sq^{L_{0}-c/24}\bar{q}^{\bar{L}_{0}-c/24}\right),\qquad Z_{\mathbb{Z}_{2}}(\tau,\bar{\tau})=\mathrm{Tr}_{\mathcal{H}_{\mathbb{Z}_{2}}}\left(q^{L_{0}-c/24}\bar{q}^{\bar{L}_{0}-c/24}\right),
\]
related by the operator that implements the $\mathbb{Z}_{2}$-symmetry.
The first is the twisted trace in the ordinary Hilbert space. The
second is the partition function in the $\mathbb{Z}_{2}$-twisted
Hilbert space. To analyze the large-$L_{z}$ limit, it is useful to
exchange the two torus cycles. In the exchanged channel, the measured
free energy difference is identified with the lowest gap in the $\mathbb{Z}_{2}$-twisted
Hilbert space. The modular $S$ transformation exchanges the two torus
cycles, relating the amplitudes, 
\[
Z_{\mathbb{Z}_{2}}(\tau,\bar{\tau})=Z^{(\mathbb{Z}_{2})}\left(-\frac{1}{\tau},-\frac{1}{\bar{\tau}}\right).
\]
Thus, inserting the symmetry line along one cycle is equivalent to
quantizing on the Hilbert space twisted along the other cycle.

For the Ising minimal model, $c=1/2$, and the three Virasoro primaries
have weights 
\[
\mathbf{1}:\  h=\bar{h}=0,\qquad\epsilon:\ h=\bar{h}=\frac{1}{2},\qquad\sigma:\  h=\bar{h}=\frac{1}{16},
\]
The diagonal modular-invariant partition function is 
\begin{equation}
Z_{\mathbf{1}}(\tau,\bar{\tau})=Z^{\text{(\textbf{1})}}(\tau,\bar{\tau})=|\chi_{0}|^{2}+|\chi_{1/2}|^{2}+|\chi_{1/16}|^{2},\label{eq:identity partition function}
\end{equation}
while insertion of the $\mathbb{Z}_{2}$ symmetry line gives 
\[
Z^{(\mathbb{Z}_{2})}(\tau,\bar{\tau})=|\chi_{0}|^{2}+|\chi_{1/2}|^{2}-|\chi_{1/16}|^{2}.
\]
The minus sign is due to the odd $\mathbb{Z}_{2}$ charge of the spin
field.

Under the modular $S$ transformation, the characters mix as 
\[
\chi_{i}\left(-\frac{1}{\tau}\right)=\sum_{j}S_{ij}\chi_{j}(\tau)\qquad\text{with}\qquad S=\frac{1}{2}\begin{pmatrix}1 & 1 & \sqrt{2}\\
1 & 1 & -\sqrt{2}\\
\sqrt{2} & -\sqrt{2} & 0
\end{pmatrix},
\]
where the basis is $(\mathbf{1},\epsilon,\sigma)$. This transformation
keeps $Z_{\boldsymbol{1}}$ invariant, up to $\tau\to-1/\tau$ and
acting on $Z^{(\mathbb{Z}_{2})}$ gives
\begin{equation}
Z_{\mathbb{Z}_{2}}(\tau,\bar{\tau})=\chi_{0}\bar{\chi}_{1/2}+\chi_{1/2}\bar{\chi}_{0}+|\chi_{1/16}|^{2}.\label{eq:z2 twisted hilbert spac}
\end{equation}
The lightest state in $\mathcal{H}_{\mathbb{Z}_{2}}$ is therefore
the $\mu$ primary, with 
\[
(h^{\mathbb{Z}_{2}}_{0},\bar{h}^{\mathbb{Z}_{2}}_{0})=\left(\frac{1}{16},\frac{1}{16}\right).
\]
Since the untwisted vacuum has $h^{\mathbf{1}}_{0}=\bar{h}^{\mathbf{1}}_{0}=0$,
the free energy density is 
\[
\Delta f(L_{z})=E^{\mathbb{Z}_{2}}_{0}-E^{\mathbf{1}}_{0}=\frac{\pi}{4L_{z}},
\]
 where we used that for a state with conformal weights $(h,\bar{h})$,
the general cylinder energy is 
\[
E(h,\bar{h})=\frac{2\pi}{L_{z}}\left(h+\bar{h}-\frac{c}{12}\right).
\]

\subsection{Finite-volume corrections }

\label{subsec:Finite-volume-corrections}

\subsubsection{Lattice corrections and irrelevant operators}

At finite-$L_{z}$, lattice realizations receive corrections from
irrelevant operators allowed by the microscopic symmetries and the
$\mathbb{Z}_{2}$-twisted ensemble. In two dimensions this structure
is simple: the critical Ising theory has only the Virasoro primaries
$\mathbf{1},\sigma,\epsilon$, so critical corrections come from irrelevant
operators in their conformal families.

For a bulk perturbation by an operator of scaling dimension $\Delta_{i}$,
\[
S_{{\rm lat}}=S_{{\rm CFT}}+\lambda_{i}\int d^{2}x\mathcal{O}_{i}(x)
\]
the correction to the cylinder energy density scales as 
\[
\delta\Delta f_{i}(L_{z})\sim L^{1-\Delta_{i}}_{z}.
\]
Equivalently, the correction to the dimensionless free energy difference
is 
\[
\delta\left[L_{z}\Delta f(L_{z})\right]\sim L^{-(\Delta_{i}-2)}_{z}.
\]

The $\mathbb{Z}_{2}$-twisted sector is equivalent to the insertion
of a topological $\mathbb{Z}_{2}$ defect line. The line position
is not physical, since it can be moved by deforming the seam. Moreover,
the ensemble selects a torus cycle but not its orientation. Since
global $\mathbb{Z}_{2}$ symmetry is preserved, $\mathbb{Z}_{2}$-odd
insertions are absent from the free energy difference. In particular,
\[
\left\langle \int d^{2}x\sigma(x)\right\rangle _{{\rm AP}}=\left\langle \int d^{2}x\sigma(x)\right\rangle _{{\rm P}}=0
\]

A nonzero one-point profile of $\sigma$ requires a pinned interface,
an oriented interface, or an explicitly symmetry-breaking field. That
is not the unpinned AP partition function used here. Reflection-odd
or odd-spin insertions are excluded for the same reason.

The energy operator is the leading $\mathbb{Z}_{2}$-even and symmetry-allowed.
It is the relevant thermal perturbation, 
\[
S_\text{pert}=t\int d^{2}x\epsilon(x),\qquad\Delta_{\epsilon}=1.
\]
It measures detuning from criticality, not an irrelevant finite-size
correction. This effect is not expected here because the critical
coupling of the two-dimensional Ising model is exact.

The leading irrelevant operators are dimension-four fields in the
identity family. A convenient basis is 
\[
T\bar{T},\qquad T^{2}+\bar{T}^{2},
\]
with the usual quasi-primary subtractions in the chiral level-four
operators understood. The operator $T\bar{T}$ is a rotational scalar.
The combination $T^{2}+\bar{T}^{2}$ has spin four and it is not invariant
under continuous rotations, but only under the $\pi/2$ rotations
of the square lattice. Both operators have $\Delta=4$ and generate
corrections of the form
\[
L_{z}\Delta f(L_{z})=\frac{\pi}{4}+\frac{a_{2}}{L^{2}_{z}}+\frac{a_{4}}{L^{4}_{z}}+\cdots.
\]
The coefficient $a_{2}$ is nonuniversal. It can receive contributions
from the scalar perturbation $T\bar{T}$ and from the lattice anisotropy
$T^{2}+\bar{T}^{2}$, unless one coefficient vanishes by improvement,
integrability, or cancellation. 

\subsubsection{Exponential corrections from the cylinder spectrum}

The leading finite-$L$ corrections can also be written explicitly.
These corrections are controlled by the spectrum of the CFT on a circle
of circumference $L_{z}$ 
\[
Z_{s}=g_{0,s}e^{-LE_{0,s}(L_{z})}\left[1+\sum_{i}r_{i,s}e^{-L\left(E_{i,s}(L_{z})-E_{0,s}(L_{z})\right)}\right],
\]
where $E_{i,s}$ are the excited state energies with $E_{0,s}$ the
ground state, $g_{0,s}$ is the ground state degeneracy, and $r_{i,s}$
are relative degeneracies, $g_{i,s}/g_{0,s}$, all in the sector $s$.
The exponential corrections are controlled by the gaps 
\[
m_{i,s}=E_{i,s}-E_{0,s}.
\]

Since, under radial quantization $\mathbb{R}^{2}\sim\mathbb{R}\times S^{1},$
the thermal, or screening, masses of the theory compactified on the
circle of size $L_{z}$ are related to the 2d Ising CFT data
\[
\kappa_{i,s}/L_{z}\equiv m_{i,s}=2\pi\left(\Delta_{i,s}-\Delta_{0,s}\right)/L_{z}.
\]
The finite-$L$ corrections are therefore 
\[
L_{z}\Delta f\left(q\right)=2\pi\left(\Delta_{0,\mathbb{Z}_{2}}-\Delta_{0,\mathbf{1}}\right)+\frac{1}{q}\left(\log\frac{1+\sum_{i}r_{i\text{,\textbf{1}}}e^{-\kappa_{i,\mathbf{1}}q}}{1+\sum_{i}r_{i,\mathbb{Z}_{2}}e^{-\kappa_{i,\mathbb{Z}_{2}}q}}-\log\left(\dfrac{g_{0,\mathbb{Z}_{2}}}{g_{0,\boldsymbol{1}}}\right)\right).
\]

The first excitation in the periodic sector is the spin primary, with
$\Delta_{\sigma}=h_{\sigma}+\bar{h}_{\sigma}=\frac{1}{8},$ such that
the leading periodic sector gap is 
\begin{equation}
\kappa_{1,\boldsymbol{1}}=2\pi\left(\Delta_{\sigma,\mathbf{1}}-\Delta_{0,\mathbf{1}}\right)=\frac{\pi}{4},\qquad r_{1,\boldsymbol{1}}=1.
\end{equation}
In the $\mathbb{Z}_{2}$-twisted Hilbert space, eq. \eqref{eq:z2 twisted hilbert spac},
the ground state is $\sigma$, with $\Delta_{0,\mathbb{Z}_{2}}=\frac{1}{8},$
and the first excited state has scaling dimension $\Delta_{1,\mathbb{Z}_{2}}=1/2$
and has relative degeneracy $r_{1,\mathbb{Z}_{2}}=2$. The corresponding
gap is 
\begin{equation}
\kappa^{\mathbb{Z}_{2}}_{1}=\frac{3\pi}{4}.
\end{equation}

Defining $t\equiv e^{-\pi q/4}$, the large-$q$ form is 
\begin{equation}
L_{z}\Delta f(q)=\frac{\pi}{4}+\frac{1}{q}\left[\log(1+t+\cdots)-\log(1+2t^{3}+\cdots)\right]\sim\frac{\pi}{4}+\frac{1}{q}\left(t-\frac{1}{2}t^{2}-\frac{5}{3}t^{3}\right)+\mathcal{O}\left(q^{-1}t^{4}\right).
\end{equation}

\subsubsection{Mixed terms}

Combining both limits, the double asymptotic expansion is
\begin{equation}
L_{z}\Delta f(L_{z},L)=\frac{\pi}{4}+\frac{a_{2}}{L^{2}_{z}}+\frac{a_{4}}{L^{4}_{z}}+\cdots+\frac{1}{q}\left[t-\frac{1}{2}t^{2}-\frac{5}{3}t^{3}-\frac{1}{4}t^{4}+\cdots\right]+\delta_{{\rm mix}}(L_{z},q).
\end{equation}
The mixed terms come from shifts of the finite-size gaps and coefficients
induced by irrelevant perturbations
\begin{equation}
\kappa_{n,s}(L_{z})=\kappa_{n,s}+\frac{\mu_{2,n,s}}{L^{2}_{z}}+\frac{\mu_{4,n,s}}{L^{4}_{z}}+\cdots,
\end{equation}
so that
\begin{equation}
\frac{1}{q}e^{-\kappa_{n,s}(L_{z})q}=e^{-\kappa_{n,s}q}\left(\frac{1}{q}-\frac{\mu_{2,n,s}}{L^{2}_{z}}+\frac{1}{L^{4}_{z}}\left(\frac{\mu^{2}_{2,n,s}q}{2}-\mu_{4,n,s}\right)+\cdots\right),
\end{equation}
and the leading mixed corrections have the schematic form 
\begin{equation}
\delta_{{\rm mix}}(L_{z},q)=\frac{e^{-\kappa_{n,s}q}}{L^{2}_{z}}\left[-\mu_{2,n,s}+\frac{1}{L^{2}_{z}}\left(\frac{\mu^{2}_{2,n,s}q}{2}-\mu_{4,n,s}\right)+\mathcal{O}(L^{-4}_{z})\right].
\end{equation}
These mixed terms are subleading unless the fit includes both intermediate
$L_{z}$ and intermediate $q$ data, but they are allowed by the same
irrelevant operators that generate the power corrections.

\section{Three-dimensional correction terms }

\label{app:3d_corrections}

This appendix generalizes to three dimensions the correction terms used in the two-dimensional benchmark. We restrict
the discussion to the leading effects relevant for the fits. One-particle
states in the transverse channel give the large-$q$ exponential corrections,
while irrelevant operators and analytic slab terms give the power-law
corrections used in the fixed-aspect ratio extrapolation.

\subsection{One-particle transverse corrections }

\label{subsec:one_particle_3d}

Quantize along one transverse direction of the box $L\times L\times L_{z}$,
and denote the sector by $s={\rm P},{\rm AP}$. The partition function
has the spectral form 
\[
Z_{s}=\mathrm{Tr}\,e^{-LH_{s}(L,L_{z})}=e^{-LE_{0,s}(L,L_{z})}g_{0,s}\left[1+\sum_{k,n,p}r_{k,s}e^{-L\sqrt{m^{2}_{k,s}+p^{2}+(2\pi n/L_{z})^{2}}}+\cdots\right],
\]
where $r_{k,s}=g_{k,s}/g_{0,s}$. For large transverse size, the sum
over momentum $p$ in the second transverse direction can be replaced
by an integral. Defining 
\[
M^{2}_{k,n,s}\equiv m^{2}_{k,s}+\left(\frac{2\pi n}{L_{z}}\right)^{2},
\]
the one-particle corrections are
\begin{equation}
Z_{s}=e^{-LE_{0,s}(L,L_{z})}g_{0,s}\left[1+\frac{L}{\pi}\sum_{k,n}r_{k,s}M_{k,n,s}K_{1}(LM_{k,n,s})+\cdots\right].\label{eq:partition function}
\end{equation}
At criticality the screening masses scale as $m_{k,s}=\kappa_{k,s}/L_{z}$.
Setting $q=L/L_{z}$ gives 
\[
M_{k,n,s}K_{1}(LM_{k,n,s})\sim\frac{1}{L_{z}}\sqrt{\frac{\pi}{2q}}\,K^{1/2}_{k,n,s}e^{-qK_{k,n,s}},\qquad K^{2}_{k,n,s}=\kappa^{2}_{k,s}+4\pi^{2}n^{2}.
\]
Thus one-particle states generate exponential corrections in $q$.
Keeping only the lightest channel in each sector, 
\[
F_{s}=LE_{0,s}(L,L_{z})-\log g_{0,s}-r_{1,s}q^{1/2}e^{-\kappa_{s}q}+\cdots.
\]

Because the transverse box is square, the finite-$L$ correction to
the ground state energy gives a second contribution. Write 
\[
E_{0,s}(L,L_{z})=E_{0,s}-\frac{1}{\pi}\sum_{k,n}g_{k,s}M_{k,n,s}K_{1}(LM_{k,n,s})+\cdots,
\]
from expanding eq. \eqref{eq:partition function} at large volume
at fixed ratio. Thus, the full correction is 
\[
F_{s}=LE_{0,s}-\log g_{0,s}-\frac{L}{\pi}\sum_{k,n}\left(g_{k,s}+r_{k,s}\right)M_{k,n,s}K_{1}(LM_{k,n,s})+\cdots.
\]
and the dimensionless free energy density becomes
\[
F_{s}(q)=C_{s}-\frac{\log g_{0,s}}{q^{2}}-\left(g_{1,s}+r_{1,s}\right)\sqrt{\dfrac{\kappa_{s}}{2\pi q^{3}}}e^{-\kappa_{s}q}+\cdots.
\]
 and the sector difference 
\begin{align*}
Y(q) & =\Delta_{\mathbb{Z}_{2}}-\left(g_{1,\text{AP}}+r_{1,\text{AP}}\right)\sqrt{\dfrac{\kappa_{\text{AP}}}{2\pi q^{3}}}e^{-q\kappa_{\text{AP}}}+\left(g_{1,\text{P}}+r_{1,\text{P}}\right)\sqrt{\dfrac{\kappa_{\text{P}}}{2\pi q^{3}}}e^{-q\kappa_{\text{P}}}-\dfrac{1}{q^{2}}\log\left(\dfrac{g_{0,\text{AP}}}{g_{0,\text{P}}}\right).
\end{align*}

\subsection{Slab power corrections }

\label{subsec:app:slab_power_corrections}

At fixed $L_{z}$, the limit $q=L/L_{z}\to\infty$ defines the slab
free energy density. As discussed in sec. \ref{sec:theory}, its large-$L_{z}$
expansion is 
\begin{equation}
Y(L_{z})=\Delta_{\mathbb{Z}_{2}}+\sum_{i}a_{i}L^{-\omega_{i}}_{z}+\sum_{n\geq1}b_{n}L^{-2n}_{z}+\cdots.\label{eq:slab_power_expansion}
\end{equation}
The first sum contains corrections from irrelevant perturbations of
the critical theory. The second contains analytic slab corrections.
We will now detail what can appear in these sums. 

The first sum follows from perturbing the three-dimensional Ising
CFT by local irrelevant operators, 
\begin{equation}
S_{{\rm lat}}=S_{{\rm CFT}}+\sum_{i}\lambda_{i}\int d^{3}x\,{\cal O}_{i}(x).
\end{equation}
At first order, only operators whose integrated insertion is invariant
under the symmetries of the AP-P observable can contribute. The unpinned
AP and P partition functions preserve the global Ising spin flip.
Thus $\mathbb{Z}_{2}$-odd bulk primaries, such as $\sigma$, $\sigma'$,
and odd tensor primaries, are absent from corrections to $F_{{\rm AP}}-F_{{\rm P}}$.
Moreover, since the 1pt functions are integrated, descendants are
projected out.

The leading $\mathbb{Z}_{2}$-even scalar irrelevant perturbation
is 
\begin{equation}
\epsilon^{\prime},\qquad\omega_{\epsilon^{\prime}}=\Delta_{\epsilon^{\prime}}-3\simeq0.82951(61),
\end{equation}
from \cite{Reehorst:2021hmp}. This is the leading ordinary bulk
correction to scaling. The relevant thermal scalar $\epsilon$ is
also symmetry-allowed, but its coefficient is the detuning from the
critical coupling, and should be absent given the precision with which
$\beta_{c}$ is known. 

Spinful primaries can contribute only if the geometry supplies an
invariant tensor with the same spin. In the present slab geometry, the compact direction, equivalently the normal to the transverse defect surface, selects an axis but not an orientation. The background is invariant under $n_{\mu}\to-n_{\mu}$.
Therefore only tensors even in $n_{\mu}$ are available. In particular,
there is no invariant vector with which to contract a spin-one primary,
so spin-one one-point functions are forbidden, unless an additional
reflection-odd structure is introduced\footnote{This exclusion is also not expected to hide a low-lying correction.
There is no known $\mathbb{Z}_{2}$-even spin-one primary in the three-dimensional
Ising CFT. If such a primary exists, existing bounds place it at $\Delta>5$
\cite{Meneses:2018xpu}. Moreover, it is not clear that the lightest
$\mathbb{Z}_{2}$-even vector, if present, would have the reflection
parity required to couple to an oriented defect structure.}.

The first allowed tensor structure is the traceless spin-two combination
\[
Q_{\mu\nu}=n_{\mu}n_{\nu}-\frac{1}{3}\delta_{\mu\nu}.
\]
The leading nonconserved \(\mathbb Z_2\)-even spin-two primary can therefore enter
the slab finite-size expansion through the scalar contraction
\[
Q^{\mu\nu}T^{\prime}_{\mu\nu},\qquad\omega_{T^{\prime}}=\Delta_{T^{\prime}}-3\simeq2.50915(44).
\]
This contribution is allowed by the slab geometry and by the symmetries of the
AP$-$P observable. The leading even spin-4 primary gives another
possible contribution, 
\begin{equation}
C_{\mu\nu\rho\sigma},\qquad\omega_{C}=\Delta_{C}-3\simeq2.022665(28),
\end{equation}
from \cite{Simmons-Duffin:2016wlq}. One source of the spin-four
structure is geometric. Both the antiperiodic direction and the rectangular
slab distinguish the compact axis. The second is the cubic lattice,
whose microscopic anisotropy supplies a spin-four tensor even when
the continuum slab geometry is rotationally symmetric in the transverse
plane.

\subsection[Sector difference of Z2-even one-point functions]{Sector difference of $\mathbb{Z}_{2}$-even one-point functions
}\label{par:Difference-one-pt torus}

We now discuss why some one-point functions may be smaller than expected. 

The one-point functions of $\mathbb{Z}_{2}$-even operators need not
be identical in the periodic and antiperiodic sectors. Nonetheless,
they may be arbitrarily close. Consider the theory on $S^{2}_{L}\times S^{1}_{L_{z}}$
and quantize on $S^{2}_{L}$. This differs from the transverse $T^{2}$
used in the simulations, but it illustrates why sector differences
can be smaller than the corresponding one-point functions in the individual
sectors.

The periodic partition function is 
\begin{equation}
Z_{{\rm P}}={\rm Tr}\,e^{-L_{z}H},
\end{equation}
whereas the antiperiodic partition function inserts the Ising spin-flip
operator $S$, 
\begin{equation}
Z_{{\rm AP}}={\rm Tr}\,Se^{-L_{z}H}.
\end{equation}
For a scalar operator ${\cal O}$, 
\begin{equation}
\langle{\cal O}\rangle_{{\rm P}}=\frac{1}{Z_{{\rm P}}}{\rm Tr}\left({\cal O}e^{-L_{z}H}\right),\qquad\langle{\cal O}\rangle_{{\rm AP}}=\frac{1}{Z_{{\rm AP}}}{\rm Tr}\left({\cal O}Se^{-L_{z}H}\right).
\end{equation}
Choose energy eigenstates that simultaneously diagonalize $S$, $S|i\rangle=\eta_{i}|i\rangle$,
with $\eta_{i}=\pm1$. Using \(E_i=\Delta_i/L\) and defining the dimensionless cylinder matrix element
\(C_{Oii}\equiv L^{\Delta_O}\langle i|O_{\rm cyl}|i\rangle\) gives \footnote{For a spinless primary state \(|i\rangle\), and with the standard normalization of
two-point functions, \(C_{Oii}\) is equal to the corresponding flat-space three-point
coefficient. More generally, in the thermal trace \(i\) can label descendants or states in
degenerate multiplets; in that case \(C_{Oii}\) should be understood as the
appropriate dimensionless cylinder matrix element, with any angular or tensor structure
and degeneracy included in this notation.}
\begin{equation}
L^{\Delta_{{\cal O}}}\langle{\cal O}\rangle_{{\rm P}}=\frac{\sum_{i}e^{-\Delta_{i}/q}C_{{\cal O}ii}}{\sum_{i}e^{-\Delta_{i}/q}},\qquad L^{\Delta_{{\cal O}}}\langle{\cal O}\rangle_{{\rm AP}}=\frac{\sum_{i}\eta_{i}e^{-\Delta_{i}/q}C_{{\cal O}ii}}{\sum_{i}\eta_{i}e^{-\Delta_{i}/q}},
\end{equation}
Therefore 
\begin{equation}
L^{\Delta_{{\cal O}}}\left(\langle{\cal O}\rangle_{{\rm P}}-\langle{\cal O}\rangle_{{\rm AP}}\right)=\sum_{i}e^{-\Delta_{i}/q}C_{{\cal O}ii}\left(\frac{1}{Z_{{\rm P}}}-\frac{\eta_{i}}{Z_{{\rm AP}}}\right).
\end{equation}

In the small-$q$ limit the lowest states dominate. Assuming a unique
$\mathbb{Z}_{2}$-even vacuum with $\Delta_{0}=0$, define 
\begin{equation}
{\cal E}(q)=\sum_{\eta_{j}=+1,\ j\neq0}e^{-\Delta_{j}/q},\qquad\Sigma(q)=\sum_{\eta_{j}=-1}e^{-\Delta_{j}/q}.
\end{equation}
Then 
\begin{equation}
Z_{{\rm P}}=1+{\cal E}+\Sigma,\qquad Z_{{\rm AP}}=1+{\cal E}-\Sigma,
\end{equation}
and 
\begin{equation}
\frac{1}{Z_{{\rm P}}}-\frac{\eta_{i}}{Z_{{\rm AP}}}=\begin{cases}
-2\Sigma+\cdots, & \eta_{i}=+1,\\[3pt]
2-2{\cal E}+\cdots, & \eta_{i}=-1.
\end{cases}
\end{equation}

In this limit, the vacuum contribution is common to the two traces
and cancels in the sector difference. Keeping only the lightest odd
state, $\sigma$, gives 
\begin{equation}
L^{\Delta_{{\cal O}}}\left(\langle{\cal O}\rangle_{{\rm P}}-\langle{\cal O}\rangle_{{\rm AP}}\right)=2g_{\sigma}C_{{\cal O}\sigma\sigma}e^{-\Delta_{\sigma}/q}+\mathcal{O}\left(e^{-\text{min}\left(2\Delta_{\sigma},\Delta_{\sigma^{\prime}}\right)/q}\right),
\end{equation}
where $g_{\sigma}$ is the degeneracy of the lightest odd level. Thus
the scalar sector difference can be much smaller than the one-point
coefficient in either sector. This supports the observation that $\epsilon^{\prime}$
corrections in the slab analysis may be numerically smaller than further
subleading corrections. 

%
%
%
%

\bibliography{PaperCasimir,main}
\bibliographystyle{JHEP}

\end{document}